\documentclass[aps,prb,amsmath,amssymb,showkeys]{revtex4}

\usepackage{graphicx}
\usepackage{subfig}
\usepackage{bm}
\usepackage{bbold}
\usepackage{hhline}

\def\br{ \bm{r} }
\def\bk{ \bm{k} }

\def\bp{ \bm{p} }
\def\spup{ \lvert\uparrow\rangle }
\def\spdown{ \lvert\downarrow\rangle }
\def\htv{ \hat{\bm{v}} }
\def\be{ \bm{e} }
\def\bH{ \bm{H} }
\def\bgam{ \bm{\gamma} }
\def\hbx{ \hat{\bm{x}} }
\def\hby{ \hat{\bm{y}} }
\def\hbz{ \hat{\bm{z}} }
\def\hcH{ \hat{\cal H} }
\def\bLam{ \bm{\Lambda} }
\def\bbell{ \bm{\ell} }

\def\tildeps{ \tilde\epsilon }
\def\diag{ \mathrm{diag} }
\def\Pf{ \mathrm{Pf}\, }

\def\im{ \mathrm{Im}\, }
\def\re{ \mathrm{Re}\, }
\def\sign{ \mathrm{sign} }

\begin{document}
\title{On the effective models of spin-orbit coupling in a two-dimensional electron gas}

\author{K. V. Samokhin\footnote{E-mail: kirill.samokhin@brocku.ca}}
\affiliation{Department of Physics, Brock University, St. Catharines, Ontario L2S 3A1, Canada}


\begin{abstract}
We use the method of invariants to derive one- and two-band effective Hamiltonians of a noncentrosymmetric two-dimensional electron gas, in the presence of magnetic field. A complete classification of the
antisymmetric spin-orbit and magnetic coupling terms near the $\Gamma$ point is developed for all two-dimensional crystal symmetries. 
The effective Hamiltonian depends on the symmetry of the Bloch bands at the $\Gamma$ point, which is described by one of the double-valued corepresentations of the two-dimensional magnetic point group. In some bands,
the spin-orbit coupling is cubic in the electron momentum and the effective Zeeman interaction is strongly anisotropic. As an example of a two-band effective Hamiltonian, we introduce a simple model of a topological 
insulator with the intraband and interband spin-orbit coupling and investigate its bulk and boundary properties.
\end{abstract}


\maketitle

\section{Introduction}
\label{sec: Intro}

Two-dimensional (2D) electron systems with the spin-orbit (SO) coupling have been one of most popular fields of research in recent years, driven primarily by their applications 
to spintronics\cite{spintronics-review-1,spintronics-review-2}
and topological superconductivity.\cite{2D-SC-review}
The long list of materials of interest includes graphene,\cite{graphene-review} monolayer transition-metal dichalcogenides,\cite{TMDC-review} 
conducting interfaces or surfaces of insulating oxides,\cite{interface-SC} ultra-thin metal films on various substrates,\cite{metal-films} and others. 
In most of these systems the inversion symmetry is broken due to the presence of a substrate, or due to different nature of the materials sandwiching the conducting layer. 

The SO coupling in a noncentrosymmetric crystal lifts the spin degeneracy of the electron states almost everywhere in the Brillouin zone, producing the Bloch bands characterized by a complex spin texture and a 
nontrivial momentum-space topology. This is responsible for a number of remarkable effects, both in normal and superconducting states, such as the quantum spin Hall effect in topological insulators,\cite{HK10} the 
spin field-effect transistor,\cite{DD90} the magnetoelectric effect,\cite{ME-effect} and the unusual nonuniform superconducting states in the presence of magnetic field or even without any field.\cite{helical-states} 

Theoretical understanding of the SO coupling-controlled physics in noncentrosymmetric materials is based on the Rashba model, see Ref. \onlinecite{Rashba-model} and also Ref. \onlinecite{Rashba-model-review} for a review. 
In the Rashba model and its generalizations, the SO coupling is described by the terms in the effective Bloch Hamiltonian which are odd in the electron momentum (in the original model,\cite{Rashba-model} 
the SO coupling is linear in $\bk$). Such terms, known as the ``asymmetric'' or ``antisymmetric'' SO coupling, are allowed by symmetry and can be justified microscopically using 
the $\bk\cdot\bp$ perturbation theory.\cite{Kittel-book} While there exists an extensive literature focusing on the crystal structures and geometries relevant to semiconducting systems, see, e.g., 
Ref. \onlinecite{Winkler-book}, a complete classification of the phenomenological effective Hamiltonians in noncentrosymmetric 2D crystals,
which includes both the SO coupling and the magnetic field interactions, seems to be missing. The goal of the present paper is to fill this gap. 

The structure of the paper is as follows. In Sec. \ref{sec: Rashba model}, we discuss the standard Rashba model and its different interpretations. In Secs. \ref{sec: Gamma-point}-\ref{sec: two-band H}, 
we use the method of invariants to derive the one-band and two-band effective Hamiltonians in the vicinity of the $\Gamma$ point. This is done for all crystal symmetries in 2D, taking into account an external magnetic field. 
We show that the structure of the effective Hamiltonian crucially depends on the $\Gamma$-point corepresentation of the 2D magnetic point group and in many
cases differs significantly from the standard Rashba form. Our investigation of two-band effective Hamiltonians is motivated by the fact that the popular models of topological insulators, such as the Kane-Mele model\cite{KM05-1}
and the Bernevig-Hughes-Zhang model,\cite{BHZ06} rely on presence of the SO coupling in a two-band, or a two-sublattice, system. In Sec. \ref{sec: TI}, 
we introduce a toy model of a 2D TR-invariant topological insulator and study its bulk and boundary properties, emphasizing the qualitatively different effects of the intraband and interband SO coupling. 
In the Appendices, we present the details of the calculations, in particular, the microscopic justification of the various terms in the phenomenological effective Hamiltonians and 
also a summary of the properties of the modified Dirac Hamiltonian. 
Throughout the paper we use the units in which $\hbar=1$, neglecting, in particular, the difference between the quasiparticle momentum and the wave vector.

\section{Rashba model}
\label{sec: Rashba model}

The electron-lattice SO coupling combined with inversion symmetry breaking lifts the twofold spin degeneracy of the Bloch states at the wave vector $\bk=(k_x,k_y)$. 
In the case of one spin-degenerate band, these effects are described by the following effective Hamiltonian, known as the Rashba model:
\begin{equation}
\label{H-Rashba}
  \hat H_{Rashba}=\epsilon(\bk)\hat\sigma_0+\bgam(\bk)\hat{\bm{\sigma}}+\mu_B\bH\hat{\bm{\sigma}},
\end{equation}
where $\hat\sigma_0$ is the unit matrix and $\hat{\bm{\sigma}}=(\hat\sigma_1,\hat\sigma_2,\hat\sigma_3)$ are the Pauli matrices. The second term is the antisymmetric SO coupling, characterized by
the pseudovector $\bgam$, which is invariant under the operations of the 2D point group, satisfies $\bgam(-\bk)=-\bgam(\bk)$, and vanishes in a centrosymmetric crystal or in the absence of the SO coupling. 
In the original Rashba model,\cite{Rashba-model} one has $\bgam(\bk)=\gamma_0(k_y\be_1-k_x\be_2)$. 
The last term in Eq. (\ref{H-Rashba}) describes the Zeeman interaction of the electron spin with a uniform external magnetic field (the spin magnetic moment is given by $-\mu_B\hat{\bm\sigma}$, where $\mu_B$ is the Bohr magneton, 
the electron charge is $-e$, and the Land\'e factor is set to $2$). The orbital effects of the magnetic field will not be considered in this paper; if needed, they can be included via the Peierls substitution,\cite{LL-9} 
whereby $\bk$ is replaced by the operator $\hat{\bk}+(e/\hbar c)\bm{A}(\br)$, where $\hat{\bk}=-i\bm{\nabla}$ and $\bm{A}$ is the vector potential. 
Given the popularity of the the Rashba model and its wide range of applications, it makes sense to revisit its microscopic justification and clarify the meaning of the various terms in Eq. (\ref{H-Rashba}). 

We consider a quasi-2D electron gas in the $xy$ plane. The potential $U(x,y,z)$ is assumed to be periodic in the $x$ and $y$ directions, but confining in the $z$ direction. Focusing on the applications
to electrons on a substrate or a surface, or in a conducting interface between two different materials, we further assume that $U(x,y,-z)\neq U(x,y,z)$, so that the system lacks inversion symmetry. 
Neglecting the electron-electron interations, the lattice vibrations, disorder, and setting $\bH=\bm{0}$, the microscopic Hamiltonian has the following form:
\begin{equation}
\label{H-general}
    \hat H=\frac{\hat{\bp}^2}{2m}+U(\br)+\frac{\hbar}{4m^2c^2}\hat{\bm{\sigma}}[\bm{\nabla}U(\br)\times\hat{\bp}],
\end{equation}
where $\hat{\bp}=-i\hbar\bm{\nabla}$ is the momentum operator and the last term describes the electron-lattice SO coupling. Regarding the symmetry of $U(\br)$, the symmetry operations (rotations and reflections) 
leaving the 2D crystal lattice invariant form the point group of the crystal, which we denote by $\mathbb{G}$. 

There are ten crystallographic point groups in 2D, also known as the rosette groups: five cyclic groups $\mathbf{C}_n$ and five dihedral groups $\mathbf{D}_n$, with $n=1$, $2$, $3$, $4$, or $6$. 
The group $\mathbf{C}_n$ has $n$ elements and is generated by the rotation $C_{nz}$ about the $z$ axis by an angle $2\pi/n$, while the group $\mathbf{D}_n$ has $2n$ elements and is generated by $C_{nz}$ and also 
by the reflection in a vertical plane (we choose $\sigma_y$ -- the reflection in the $y=0$ plane -- as the second generating element). In addition to the rosette group operations, in the absence of an external magnetic field 
the Hamiltonian (\ref{H-general}) is also invariant under the time reversal (TR) operation $K$.

There are several conceptually different ways to construct the effective models of the SO coupling in an inversion-asymmetric system, depending on how Eq. (\ref{H-general}) is partitioned to represent various physical interactions.
One possible approach is to represent the microscopic Hamiltonian in the form $\hat H=\hat H_0+\hat H_{SOC}$, where $\hat H_0$ comprises the first two terms in Eq. (\ref{H-general}), while $\hat H_{SOC}$ is the third term. 
Assuming that the SO coupling is switched on in the presence of an inversion-asymmetric crystal potential, one first finds the eigenstates of $\hat H_0$, which are twofold degenerate due to spin, 
and then calculates the matrix elements of $\hat H_{SOC}$ in this basis. In this approach, $\hat H_{Rashba}$ is a $2\times 2$ matrix in the spin space, $\epsilon(\bk)$ is the band dispersion in a noncentrosymmetric crystal 
without the SO coupling, and $\bgam(\bk)$ describes the effects of the SO coupling. Such perturbative treatment of the SO coupling might be problematic, especially in compounds containing heavy elements.

Alternatively, one can split the Hamiltonian (\ref{H-general}) into the inversion-symmetric and inversion-antisymmetric parts: $\hat H=\hat H_s+\hat H_a$, both including the SO coupling. 
The eigenstates of $\hat H_s$ are twofold degenerate due to the combined symmetry operation $KI$, called conjugation ($I$ is the space inversion operation). Then, one calculates the matrix elements of $\hat H_a$ 
in the eigenbasis of $\hat H_s$. In this approach, $\hat H_{Rashba}$ is a $2\times 2$ matrix in the conjugacy space, $\epsilon(\bk)$ is the band dispersion in a centrosymmetric lattice potential with the SO coupling, 
and $\bgam(\bk)$ describes the effects of inversion symmetry breaking. While this approach may be legitimate in three dimensional (3D) crystals with a strong SO coupling,\cite{Sam19-1} the possibility to treat $\hat H_a$ 
as a perturbation in a 2D electron gas on a substrate is questionable.  

In this paper, we use another approach, which is based on the method of invariants in the $\bk\cdot\bm{p}$ perturbation theory.\cite{Lutt56,Bir-Pikus-book} In this approach, one starts with the exact basis of the Bloch states 
at the $\Gamma$ point at zero magnetic field, which includes both the SO coupling \textit{and} the inversion symmetry breaking, and then constructs the effective Hamiltonian using an expansion in powers of $\bk$ and $\bH$
in the vicinity of the $\Gamma$ point. Assuming that the Bloch states at the $\Gamma$ point are $d_\Gamma$-fold degenerate, the effective Hamiltonian is given by a $d_\Gamma\times d_\Gamma$ matrix. 
We shall see that in our 2D case, $d_\Gamma$ is always equal to two, however, the effective Hamiltonian does not necessarily have the simple form (\ref{H-Rashba}).

\section{Bloch states at the $\Gamma$ point}
\label{sec: Gamma-point}

Let us first look in detail at the symmetry of the the Bloch states at $\bk=\bm{0}$ in the absence of magnetic field. The reduced Hamiltonian at the $\Gamma$ point has the same form as Eq. (\ref{H-general}) 
and its full symmetry group is given by
\begin{equation}
\label{magnetic G}
  {\cal G}_{\bk=\bm{0}}=\mathbb{G}+K\mathbb{G},
\end{equation}
where the TR operation $K$ commutes with all elements of the rosette group $\mathbb{G}$. Since $K$ is antiunitary, ${\cal G}_{\bk=\bm{0}}$ is a Type II magnetic, or Shubnikov, point group, see, e.g., Ref. \onlinecite{BC-book}.
At other TR invariant momenta, which satisfy $-\bk=\bk+\bm{G}$, where $\bm{G}$ is a reciprocal lattice vector, the full symmetry group also contains $K$ and has a form similar to Eq. (\ref{magnetic G}), with 
the rosette group replaced by one of its subgroups.

The Bloch states at $\bk=\bm{0}$ transform according to the irredicible double-valued corepresentations (coreps) of the magnetic group (\ref{magnetic G}). The coreps can be obtained from the double-valued 
irreducible representations (irreps) of the unitary component $\mathbb{G}$ using a standard procedure\cite{BD68,BC-book} and are classified into three cases, A, B, or C, 
which determine whether or not the antiunitary symmetry brings about an additional degeneracy and also the type of this degeneracy. While there is no additional degeneracy due to TR in Case A 
(i.e. Case A coreps are the same as the usual irreps), in the other two cases the coreps have twice the dimension of the irreps they are constructed from, leading to an additional degeneracy of the ``doubling'' type 
in Case B or the ``pairing'' type in Case C. In the latter case, the coreps are obtained by combining two complex-conjugate irreps.

Since each rosette group is isomorphic to a certain 3D point group, its double-valued coreps can be easily found, with the results listed in Table \ref{table: coreps-point} (our notations for the irreps 
are the same as in Ref. \onlinecite{Lax-book}). Due to the absence of inversion symmetry, the coreps do not have a definite parity. 
It is important to note that all double-valued coreps of the rosette groups are 2D, therefore the electron bands are twofold degenerate at the $\Gamma$ point. 
The Bloch states $|\bk=\bm{0},n,s\rangle\equiv|n,s\rangle$ are labelled by the band index $n$ and the additional index $s=1,2$, which distinguishes two orthonormal states within the same band:  
\begin{equation}
\label{Bloch-states-Gamma-point}
  |n,1\rangle,\quad |n,2\rangle=K|n,1\rangle.
\end{equation}
Due to the presence of the SO coupling, the Kramers index $s$ is not the same as the electron spin projection. 
If the states $|n,1\rangle$ and $|n,2\rangle$ form the basis of a corep described by $2\times 2$ matrices $\hat{\cal D}_n(g)$, where $g\in\mathbb{G}$, then we have
\begin{equation}
\label{basis-transform-g}
    g|n,s\rangle=\sum_{s'}|n,s'\rangle {\cal D}_{n,s's}(g).
\end{equation}
Using also $\hat{\cal D}_n(K)=-i\hat\sigma_2$, one can obtain the corep matrices for all elements of the magnetic group (\ref{magnetic G}).

\begin{table}
\caption{The double-valued coreps of the rosette groups at the $\Gamma$ point. Each rosette group is isomorphic to some noncentrosymmetric 3D point group $\mathbb{G}_{3D}$. 
All coreps are 2D. The Case C coreps are obtained by pairing two complex-conjugate 1D irreps. The last column shows whether the Bloch states at the $\Gamma$ point transform as the basis spinors.}
\begin{tabular}{|c|c|c|c|c|}
    \hline
    $\quad \mathbb{G}\quad $ & $\quad\mathbb{G}_{3D}\quad$ & corep &  \ corep case\ \ & \ pseudospin\ \ \\ \hline
    $\mathbf{C}_{1}$ & $\mathbf{C}_{1}$ & $\Gamma_2$ &  B & Y  \\ \hline
    $\mathbf{C}_{2}$ & $\mathbf{C}_{2}$ & $(\Gamma_3,\Gamma_4)$ &  C & Y \\ \hline
    $\mathbf{C}_{3}$ & $\mathbf{C}_{3}$ & $(\Gamma_4,\Gamma_5)$  &  C & Y  \\ 
                     &                  & $\Gamma_6$ &  B & N  \\ \hline
    $\mathbf{C}_{4}$ & $\mathbf{C}_{4}$ & $(\Gamma_5,\Gamma_6)$  &  C & Y \\ 
                     & & $(\Gamma_7,\Gamma_8)$  &  C & N  \\ \hline
    $\mathbf{C}_{6}$ & $\mathbf{C}_{6}$ & $(\Gamma_7,\Gamma_8)$  &  C & Y  \\ 
		     & & $(\Gamma_9,\Gamma_{10})$  &  C & N  \\
                     & & $(\Gamma_{11},\Gamma_{12})$  &  C & N  \\ \hline
    $\mathbf{D}_{1}$ & $\mathbf{C}_{s}$ & $(\Gamma_3,\Gamma_4)$ &  C & Y \\ \hline   
    $\mathbf{D}_{2}$ & $\mathbf{C}_{2v}$ & $\Gamma_5$  &  A & Y  \\ \hline
    $\mathbf{D}_{3}$ & $\mathbf{C}_{3v}$ & $\Gamma_4$  &  A & Y  \\ 
		     & & $(\Gamma_5,\Gamma_6)$ &  C & N  \\ \hline
    $\mathbf{D}_{4}$ & $\mathbf{C}_{4v}$ & $\Gamma_6$  &  A & Y \\ 
		     & & $\Gamma_7$  &  A & N \\ \hline 			
    $\mathbf{D}_{6}$ & $\mathbf{C}_{6v}$ & $\Gamma_7$ &  A & Y  \\ 
                     & & $\Gamma_8$ &  A & N  \\ 
                     & & $\Gamma_9$ &  A & N  \\ \hline
\end{tabular}
\label{table: coreps-point}
\end{table}

If the $\Gamma$-point corep is equivalent to the spin-$1/2$ corep, then the band is called ``pseudospin band'' and one can put
\begin{equation}
\label{Ueda-Rice}
  \hat{\cal D}_n(g)=\hat D^{(1/2)}(R)
\end{equation}
for $g=R$ or $IR$, where $\hat D^{(1/2)}(R)=e^{-i\theta(\bm{n}\hat{\bm{\sigma}})/2}$ is the spin-1/2 representation of a counterclockwise rotation $R$ through an angle $\theta$ about an axis $\bm{n}$. 
In general, however, the Bloch states at the $\Gamma$ point do not transform under the rosette group operations in the same way as the basis spinors $\spup$ and $\spdown$ (the eigenstates of the spin operator $\hat s_z$), 
which means that the $\Gamma$-point corep is not equivalent to the spin-$1/2$ corep and
$$
  \hat{\cal D}_n(g)\neq\hat D^{(1/2)}(R)
$$
for some $g$. We call such bands ``nonpseudospin bands''. The representation matrices for the nonpseudospin double-valued coreps are shown in Table \ref{table: nonpseudospin-corep-matrices}, 
see Appendix \ref{app: Bloch bases} for details. 

At a generic momentum in the 2D Brillouin zone, such that $\bk$ is invariant neither under TR nor any rosette group operations, the full symmetry group of the reduced Hamiltonian is given by ${\cal G}_{\bk}=\mathbf{C}_1$. 
This group does not contain any antiunitary elements, therefore the Bloch states are classified according to the usual double-valued irreps of $\mathbf{C}_1$. Since there is only one such irrep, namely $\Gamma_2$, 
which is one-dimensional (1D), the bands are nondegenerate at a generic $\bk$.

\begin{table}
\caption{The nonpseudospin double-valued corep matrices at the $\Gamma$ point ($g$ denotes the generators of the rosette group $\mathbb{G}$). }
\begin{tabular}{|c|c|c|}
    \hline
    $\quad \mathbb{G}\quad $ &  corep $\Gamma$ & $\hat{\cal D}_\Gamma(g)$  \\ \hline
    $\mathbf{C}_{3}$  & $\Gamma_6$ &  $\hat{\cal D}(C_{3z})=-\hat\sigma_0$  \\ \hline
    $\mathbf{C}_{4}$  & $(\Gamma_7,\Gamma_8)$   &  $\hat{\cal D}(C_{4z})=-\hat D^{(1/2)}(C_{4z})$  \\ \hline
    $\mathbf{C}_{6}$  & $(\Gamma_{9},\Gamma_{10})$  & $\hat{\cal D}(C_{6z})=-\hat D^{(1/2)}(C_{6z})$  \\ \cline{2-3}
                      & $(\Gamma_{11},\Gamma_{12})$  & $\hat{\cal D}(C_{6z})=-\hat D^{(1/2)}(C_{2z})$  \\ \hline
    $\mathbf{D}_{3}$  & $(\Gamma_5,\Gamma_6)$  & $\hat{\cal D}(C_{3z})=-\hat\sigma_0$,\quad $\hat{\cal D}(\sigma_y)=\hat D^{(1/2)}(C_{2y})$  \\ \hline								
    $\mathbf{D}_{4}$ & $\Gamma_7$  & \ $\hat{\cal D}(C_{4z})=-\hat D^{(1/2)}(C_{4z})$,\quad $\hat{\cal D}(\sigma_y)=\hat D^{(1/2)}(C_{2y})$ \\ \hline
    $\mathbf{D}_{6}$  & $\Gamma_8$  & $\hat{\cal D}(C_{6z})=-\hat D^{(1/2)}(C_{6z})$,\quad $\hat{\cal D}(\sigma_y)=\hat D^{(1/2)}(C_{2y})$  \\ \cline{2-3}
                       & $\Gamma_9$ & $\hat{\cal D}(C_{6z})=-\hat D^{(1/2)}(C_{2z})$,\quad $\hat{\cal D}(\sigma_y)=\hat D^{(1/2)}(C_{2y})$  \\ \hline                                                            
\end{tabular}
\label{table: nonpseudospin-corep-matrices}
\end{table}

\section{One-band effective Hamiltonian}
\label{sec: one-band H}

The band structure in the vicinity of the $\Gamma$ point can be described by an effective $\bk$-dependent Hamiltonian, whose eigenvalues correspond to the band energies. Including the Zeeman interaction, 
the effective Hamiltonian becomes also a function of the magnetic field $\bH$. In the one-band case, i.e., if just one $\Gamma$-point corep is taken into account, the effective Hamiltonian is given in the basis 
(\ref{Bloch-states-Gamma-point}) by a $2\times 2$ Hermitian matrix $\hcH(\bk,\bH)$, which depends on $\bk$ and $\bH$ as parameters and has the following general form (in this section we drop the band index $n$):
\begin{equation}
\label{H-eff-general}
  \hcH(\bk,\bH)=\varepsilon(\bk,\bH)\hat\sigma_0+\bLam(\bk,\bH)\hat{\bm\sigma},
\end{equation}
where the functions $\varepsilon$ and $\bLam=(\Lambda_1,\Lambda_2,\Lambda_3)$ contain the effects of the SO coupling and inversion symmetry breaking. Diagonalizing this Hamiltonian, one obtains two bands 
\begin{equation}
\label{helicity bands}
  \xi_\lambda(\bk,\bH)=\varepsilon(\bk,\bH)+\lambda|\bLam(\bk,\bH)|,
\end{equation}
labelled by the ``helicity'' index $\lambda=\pm$. The helicity bands are nondegenerate, except the points where all three components of $\bLam$ simultaneously vanish. 
In this section, we use a phenomenological, symmetry-based, approach known as the method of invariants\cite{Bir-Pikus-book} to find the form of $\varepsilon$ and $\bLam$. 
The parameters of the effective Hamiltonian can be calculated microscopically, using the $\bk\cdot\bp$ perturbation theory, see Appendix \ref{app: k p}.

Under an operation $g$ from the rosette group $\mathbb{G}$, the basis functions are transformed into $g|s\rangle=\sum_{s'}|s'\rangle {\cal D}_{s's}(g)$, see Eq. (\ref{basis-transform-g}). 
On the other hand, the matrix elements of $\hcH(\bk,\bH)$ in the ``old'' basis must be equal to the matrix elements of the effective Hamiltonian with the transformed parameters $\hcH(g\bk,g\bH)$
in the ``new'' basis. In this way, we obtain the following constraint on the matrix $\hcH$:
\begin{equation}
\label{H-constraint-g}
  \hcH(\bk,\bH)=\hat{\cal D}^\dagger(g)\hcH(g\bk,g\bH)\hat{\cal D}(g).
\end{equation}
Regarding the invariance under TR operation, the symmetry of the matrix $\hcH(\bk,\bH)$ is the same as that of the matrix representation of an observable $\hat{\cal O}$ which depends on $\bk$ and $\bH$ and 
satisfies the following condition: $K\hat{\cal O}(\bk,\bH)K^{-1}=\hat{\cal O}(-\bk,-\bH)$. Using $K|i\rangle=\sum_{i_1}|i_1\rangle{\cal D}_{i_1i}(K)$, where $\hat{\cal D}(K)$ is the representation of TR in an
arbitrary basis $|i\rangle$, and the antiunitarity property $\langle i|K^{-1}|j\rangle=\langle Ki|j\rangle^*$, we obtain for the matrix elements of the operator $\hat{\cal O}$: 
\begin{eqnarray*}
  \langle i|\hat{\cal O}(\bk,\bH)|j\rangle &=& \langle i|K^{-1}\hat{\cal O}(-\bk,-\bH)K|j\rangle \\
  &=& \sum_{j_1}{\cal D}_{j_1j}^*(K)\langle i|K^{-1}\hat{\cal O}(-\bk,-\bH)|j_1\rangle \\
  &=& \sum_{i_1j_1}{\cal D}_{i_1i}(K){\cal D}_{j_1j}^*(K)\langle i_1|\hat{\cal O}(-\bk,-\bH)|j_1\rangle^*.
\end{eqnarray*}
Therefore,
\begin{equation}
\label{H-constraint-K}
  \hcH(\bk,\bH)=\hat{\cal D}^\top(K)\hcH^*(-\bk,-\bH)\hat{\cal D}^*(K),
\end{equation}
where $\hat{\cal D}(K)=-i\hat\sigma_2$ in the basis (\ref{Bloch-states-Gamma-point}). Note that the constraints (\ref{H-constraint-g}) and (\ref{H-constraint-K}) are completely general and 
can also be applied to multiband effective Hamiltonians. If $N$ bands (coreps) are included in $\hcH$, then $\hat{\cal D}(g)$ and $\hat{\cal D}(K)$ become $2N\times 2N$ block-diagonal matrices, 
see Sec. \ref{sec: two-band H} for $N=2$. 

Substituting Eq. (\ref{H-eff-general}) in the TR invariance constraint (\ref{H-constraint-K}), we obtain
\begin{equation}
\label{TR-epsilon-Gamma}
  \varepsilon(\bk,\bH)=\varepsilon(-\bk,-\bH),\quad \bLam(\bk,\bH)=-\bLam(-\bk,-\bH),
\end{equation}
while the point-group invariance condition (\ref{H-constraint-g}) yields
\begin{equation}
\label{g-epsilon-Gamma}
  \varepsilon(\bk,\bH)=\varepsilon(g^{-1}\bk,g^{-1}\bH),\quad \bLam(\bk,\bH)=\hat{\cal R}(g)\bLam(g^{-1}\bk,g^{-1}\bH).
\end{equation}
The $3\times 3$ orthogonal matrix $\hat{\cal R}$ depends on the $\Gamma$-point corep and is defined by the following expression:
\begin{equation}
\label{sigma-rotation-nonpseudospin}
  \hat{\cal D}^\dagger(g)\hat{\sigma}_\mu\hat{\cal D}(g)=\sum_{\nu}{\cal R}_{\mu\nu}(g)\hat{\sigma}_\nu.
\end{equation}
Here and below we use the Greek indices ($\mu,\nu=1,2,3$) to label the Pauli matrices in the corep space and the corresponding components of $\bLam$, and the Latin indices ($i,j=x,y,z$) to label the components of vectors 
and pseudovectors, such as $\bk$ and $\bH$, in physical space. 
It follows from Eqs. (\ref{TR-epsilon-Gamma}) and (\ref{g-epsilon-Gamma}) that the helicity bands (\ref{helicity bands}) satisfy $\xi_\lambda(\bk,\bH)=\xi_\lambda(-\bk,-\bH)$ and 
$\xi_\lambda(\bk,\bH)=\xi_\lambda(g^{-1}\bk,g^{-1}\bH)$. When applying the rosette group invariance conditions one should keep in mind that the proper and improper symmetry elements act 
differently on the vector $\bk$ and the pseudovector $\bH$: for a proper rotation $g=R$, we have $g^{-1}\bk=R^{-1}\bk$ and $g^{-1}\bH=R^{-1}\bH$, but for an improper rotation $g=IR$, we have 
$g^{-1}\bk=-R^{-1}\bk$ and $g^{-1}\bH=R^{-1}\bH$.

In all pseudospin bands, the corep matrix has the form (\ref{Ueda-Rice}) and, using the identity
$$
  \hat{D}^{(1/2),\dagger}(R)\hat{\sigma}_\mu\hat{D}^{(1/2)}(R)=\sum_{\nu}R_{\mu\nu}\hat{\sigma}_\nu,
$$
where $\hat R$ is the $3\times 3$ rotation matrix, we find that $\hat{\cal R}(g)=\hat R(g)$ for all elements of the rosette group. This is also valid in most nonpseudospin bands, as one can check using the corep matrices 
from Table \ref{table: nonpseudospin-corep-matrices}. However, there are four exceptional coreps in trigonal and hexagonal crystals, in which $\hat{\cal R}(g)\neq\hat R(g)$ for some elements of the rosette group. 
For $\Gamma_6$ of $\mathbf{C}_{3}$, we have $\hat{\cal R}(C_{3z})=\hat{\mathbb 1}$ (the $3\times 3$ unit matrix); for $(\Gamma_{11},\Gamma_{12})$ of $\mathbf{C}_{6}$, we have $\hat{\cal R}(C_{6z})=\hat R(C_{2z})$; 
for $(\Gamma_5,\Gamma_6)$ of $\mathbf{D}_{3}$, we have $\hat{\cal R}(C_{3z})=\hat{\mathbb 1}$ and $\hat{\cal R}(\sigma_{y})=\hat R(C_{2y})$; and for $\Gamma_9$ of $\mathbf{D}_{6}$, we have $\hat{\cal R}(C_{6z})=\hat R(C_{2z})$ 
and $\hat{\cal R}(\sigma_{y})=\hat R(C_{2y})$.
We shall see below that the effective Hamiltonian in these bands has a form which is considerably different from the non-exceptional bands. 

We focus on the response of the system to a weak magnetic field and expand the effective Hamiltonian in powers of $\bH$:
\begin{equation}
\label{eps-Gam-expansion}
  \begin{array}{l} 
    \varepsilon(\bk,\bH)=\epsilon(\bk)+\bH\bbell(\bk)+O(H^2), \medskip \\ 
    \Lambda_\nu(\bk,\bH)=\gamma_\nu(\bk)+\sum\limits_i H_i\mu_{i\nu}(\bk)+O(H^2),
  \end{array}
\end{equation}
where $\epsilon$, $\bbell$, $\bgam$, and $\hat\mu$ are real functions of the wave vector. While the first term in $\bLam$ reproduces the antisymmetric SO coupling in Eq. (\ref{H-Rashba}), 
the interaction with an external magnetic field does not, in general, have the simple Zeeman form.

\subsection{Antisymmetric SO coupling}
\label{sec: ASOC}

From the requirements of TR and point-group invariance, see Eqs. (\ref{TR-epsilon-Gamma}) and (\ref{g-epsilon-Gamma}), we obtain:
\begin{equation}
\label{epsilon-invariance}
  \epsilon(-\bk)=\epsilon(\bk),\quad \epsilon(\bk)=\epsilon(g^{-1}\bk),
\end{equation}
i.e., $\epsilon$ is an invariant scalar, for which one can put $\epsilon(\bk)=k^2/2m^*$, where $m^*$ is the effective mass. For the antisymmetric SO coupling, we have
\begin{equation}
\label{gamma-invariance}
  \bgam(-\bk)=-\bgam(\bk),\quad \bgam(\bk)=\hat{\cal R}(g)\bgam(g^{-1}\bk).
\end{equation}
In the non-exceptional bands we have $\hat{\cal R}(g)=\hat R(g)$ for all $g$, therefore $\bgam$ is an invariant pseudovector.  
In contrast, in the four exceptional cases we have $\hat{\cal R}(g)\neq\hat R(g)$ for some $g$, so that $\bgam$ does not transform as an invariant pseudovector. Expressions for the antisymmetric SO coupling 
in the vicinity of the $\Gamma$ point for all rosette groups are listed in Table \ref{table: gammas} (for non-exceptional bands, these expressions are the same as in Ref. \onlinecite{Sam15}), with   
an example of the calculation given in Appendix \ref{app: D_3}. For a classification of the antisymmetric SO coupling in 3D crystals, see Refs. \onlinecite{Smidman-review} and \onlinecite{Sam19-1}.

Note that for $\mathbb{G}=\mathbf{C}_2$, $\mathbf{C}_4$, $\mathbf{C}_6$, $\mathbf{D}_2$, $\mathbf{D}_4$, and $\mathbf{D}_6$ the SO coupling is ``planar'', in the sense that $\gamma_3$ vanishes at all $\bk$. 
This symmetry-imposed constraint is due to the presence of the rotation $C_{2z}$ in these rosette groups. Indeed, putting $g=C_{2z}$ in Eq. (\ref{gamma-invariance}) and observing that 
$\hat{\cal R}(C_{2z})=\hat R(C_{2z})$ for all coreps, we obtain: $\gamma_3(\bk)=C_{2z}\gamma_3(C^{-1}_{2z}\bk)=\gamma_3(-\bk)=-\gamma_3(\bk)$, therefore $\gamma_3(\bk)=0$. 

Also note that in the exceptional bands, the linear in $\bk$ terms are not allowed by symmetry and the antisymmetric SO coupling is cubic in $\bk$. This should be contrasted with the ``cubic Rashba'' model 
discussed in the context of quasi-2D semiconductor quantum wells,\cite{Winkler-book} or surface states in rare-earth compounds,\cite{Usach20} in which the linear in momentum terms are numerically small 
due to material-specific reasons.

\subsection{Magnetic field coupling}
\label{sec: H-coupling}

The magnetic field enters two different terms in the effective Hamiltonian (\ref{eps-Gam-expansion}): the term containing $\hat\mu(\bk)$, which formally resembles the usual Zeeman coupling, 
albeit with an anisotropic magneton, and also the term containing $\bbell(\bk)$, which is absent in the standard Rashba model (\ref{H-Rashba}). 
Both terms originate from the microscopic Zeeman interaction modified by the SO coupling and can be obtained using the $\bk\cdot\bp$ perturbation theory, see Appendix \ref{app: k p}. 
From Eqs. (\ref{TR-epsilon-Gamma}) and (\ref{g-epsilon-Gamma}), we obtain the following symmetry constraints:
\begin{equation}
\label{ell-invariance}
  \bbell(-\bk)=-\bbell(\bk),\quad \bbell(\bk)=\hat R(g)\bbell(g^{-1}\bk)
\end{equation}
and
\begin{equation}
\label{mu-invariance}
  \hat\mu(-\bk)=\hat\mu(\bk),\quad \hat\mu(\bk)=\hat R(g)\hat\mu(g^{-1}\bk)\hat{\cal R}^{-1}(g)
\end{equation}
for all $g\in\mathbb{G}$.

We see that $\bbell(\bk)$ transforms an invariant pseudovector in all bands, pseudospin and nonpseudospin. It has the same structure as $\bgam(\bk)$ in the
non-exceptional bands, with the results listed in Table \ref{table: ells}. In particular, $\ell_z(\bk)=0$ if the rotation $C_{2z}$ is a symmetry element. Note that the absence of inversion symmetry is important: 
if the point group contained $g=I$, then both $\bgam$ and $\bbell$ would vanish identically, according to Eqs. (\ref{gamma-invariance}) and (\ref{ell-invariance}).

\begin{table}
\caption{Lowest-order polynomial expressions for the antisymmetric SO coupling ($a_i$ are real constants, $b_i$ are complex constants, and $k_\pm=k_x\pm ik_y$). Asterisks label the 
exceptional nonpseudospin bands, in which $\hat{\cal R}(g)\neq\hat R(g)$ for some elements of the rosette group.}
\begin{tabular}{|c|c|c|}
    \hline
    \hspace*{2mm} $\mathbb{G}$ \hspace*{2mm} & $\Gamma$ & $\bgam(\bk)$ \\ \hline
    $\mathbf{C}_1$   & $\Gamma_2$ & $(a_1k_x+a_2k_y)\be_1+(a_3k_x+a_4k_y)\be_2+(a_5k_x+a_6k_y)\be_3$  \\ \hline
    $\mathbf{C}_2$   & $(\Gamma_3,\Gamma_4)$ & $(a_1k_x+a_2k_y)\be_1+(a_3k_x+a_4k_y)\be_2$  \\ \hline
    $\mathbf{C}_3$   & $(\Gamma_4,\Gamma_5)$ & $(a_1k_x+a_2k_y)\be_1+(-a_2k_x+a_1k_y)\be_2+(b_1k_+^3+b_1^*k_-^3)\be_3$  \\ 
                     & $\Gamma_6{}^*$ & $(b_1k_+^3+b_1^*k_-^3)\be_1+(b_2k_+^3+b_2^*k_-^3)\be_2+(b_3k_+^3+b_3^*k_-^3)\be_3$ \\ \hline
    $\mathbf{C}_4$   & $(\Gamma_5,\Gamma_6)$ & $(a_1k_x+a_2k_y)\be_1+(-a_2k_x+a_1k_y)\be_2$  \\ 
                     & $(\Gamma_7,\Gamma_8)$ & $(a_1k_x+a_2k_y)\be_1+(-a_2k_x+a_1k_y)\be_2$  \\ \hline
    $\mathbf{C}_6$   & $(\Gamma_7,\Gamma_8)$ & $(a_1k_x+a_2k_y)\be_1+(-a_2k_x+a_1k_y)\be_2$  \\ 
                     & $(\Gamma_9,\Gamma_{10})$ & $(a_1k_x+a_2k_y)\be_1+(-a_2k_x+a_1k_y)\be_2$ \\ 
                     & $(\Gamma_{11},\Gamma_{12}){}^*$ & $(b_1k_+^3+b_1^*k_-^3)\be_1+(b_2k_+^3+b_2^*k_-^3)\be_2$ \\ \hline
    $\mathbf{D}_1$   & $(\Gamma_3,\Gamma_4)$ & $a_1k_y\be_1+a_2k_x\be_2+a_3k_y\be_3$  \\ \hline
    $\mathbf{D}_2$   & $\Gamma_5$ & $a_1k_y\be_1+a_2k_x\be_2$  \\ \hline
    $\mathbf{D}_3$   & $\Gamma_4$ & $a_1(k_y\be_1-k_x\be_2)+ia_2(k_+^3-k_-^3)\be_3$  \\ 
                     & $(\Gamma_5,\Gamma_6){}^*$ & $ia_1(k_+^3-k_-^3)\be_1+a_2(k_+^3+k_-^3)\be_2+ia_3(k_+^3-k_-^3)\be_3$  \\ \hline
    $\mathbf{D}_4$   & $\Gamma_6$ & $a_1(k_y\be_1-k_x\be_2)$  \\ 
                     & $\Gamma_7$ & $a_1(k_y\be_1-k_x\be_2)$  \\ \hline
    $\mathbf{D}_6$   & $\Gamma_7$ & $a_1(k_y\be_1-k_x\be_2)$  \\ 
                     & $\Gamma_8$ & $a_1(k_y\be_1-k_x\be_2)$  \\ 
                     & $\Gamma_9{}^*$ & $ia_1(k_+^3-k_-^3)\be_1+a_2(k_+^3+k_-^3)\be_2$ \\ \hline
\end{tabular}
\label{table: gammas}
\end{table}

\begin{table}
\caption{Lowest-order polynomial expressions for the magnetic coupling $\bbell(\bk)$ ($a_i$ are real constants, $b_i$ are complex constants, and $k_\pm=k_x\pm ik_y$). }
\begin{tabular}{|c|c|}
    \hline
    \hspace*{2mm} $\mathbb{G}$ \hspace*{2mm} & $\bbell(\bk)$ \\ \hline
    $\mathbf{C}_1$   & $(a_1k_x+a_2k_y)\hbx+(a_3k_x+a_4k_y)\hby+(a_5k_x+a_6k_y)\hbz$  \\ \hline
    $\mathbf{C}_2$   & $(a_1k_x+a_2k_y)\hbx+(a_3k_x+a_4k_y)\hby$  \\ \hline
    $\mathbf{C}_3$   & $(a_1k_x+a_2k_y)\hbx+(-a_2k_x+a_1k_y)\hby+(b_1k_+^3+b_1^*k_-^3)\hbz$  \\ \hline
    $\mathbf{C}_4$   & $(a_1k_x+a_2k_y)\hbx+(-a_2k_x+a_1k_y)\hby$  \\ \hline
    $\mathbf{C}_6$   & $(a_1k_x+a_2k_y)\hbx+(-a_2k_x+a_1k_y)\hby$ \\ \hline
    $\mathbf{D}_1$   & $a_1k_y\hbx+a_2k_x\hby+a_3k_y\hbz$  \\ \hline
    $\mathbf{D}_2$   & $a_1k_y\hbx+a_2k_x\hby$  \\ \hline
    $\mathbf{D}_3$   & $a_1(k_y\hbx-k_x\hby)+ia_2(k_+^3-k_-^3)\hbz$ \\ \hline
    $\mathbf{D}_4$   & $a_1(k_y\hbx-k_x\hby)$  \\ \hline
    $\mathbf{D}_6$   & $a_1(k_y\hbx-k_x\hby)$  \\ \hline
\end{tabular}
\label{table: ells}
\end{table}

It follows from Eq. (\ref{mu-invariance}) that the effective Zeeman coupling $\hat\mu(\bk)$ is an invariant second-rank tensor only in non-exceptional bands, when $\hat{\cal R}(g)=\hat R(g)$.
In these bands, we have $\hat\mu(\bk)=\hat R(g)\hat\mu(g^{-1}\bk)\hat R^{-1}(g)$, therefore the simplest expression for the effective Zeeman coupling has the usual, 
i.e., isotropic and momentum-independent, form: $\mu_{i\nu}(\bk)=\mu_m\delta_{i\nu}$. The effective ``magneton'' $\mu_m$ includes the effects of the crystal field and the SO coupling and 
does not have to be equal to the Bohr magneton $\mu_B$. If needed, one can use a more general expression
\begin{equation}
\label{usual-Zeeman}
  \hat\mu(\bk)=\left(\begin{array}{ccc}
                       \mu_\perp & 0 & 0 \\
                       0 & \mu_\perp & 0 \\
                       0 & 0 & \mu_z
                       \end{array}\right),
\end{equation}
which distinguishes between the basal-plane and $z$-axis Zeeman couplings, but is still $\bk$-independent.

In contrast, in the exceptional bands in trigonal and hexagonal crystals the effective Zeeman coupling is essentially anisotropic. In the $\Gamma_6$ bands for $\mathbb{G}=\mathbf{C}_{3}$ and in the
$(\Gamma_{11},\Gamma_{12})$ bands for $\mathbb{G}=\mathbf{C}_{6}$ we have
\begin{widetext}
\begin{equation}
\label{mu-C_3-C_6-final}
    \hat\mu(\bk)=\left(\begin{array}{ccc}
                       \beta_1k_+^2+\beta_1^*k_-^2 & \beta_2k_+^2+\beta_2^*k_-^2 & 0 \\
                       i\beta_1k_+^2-i\beta_1^*k_-^2 & i\beta_2k_+^2-i\beta_2^*k_-^2 & 0 \\
                       0 & 0 & \mu_z
                       \end{array}\right),
\end{equation}
\end{widetext}
where $\beta_{1,2}$ are complex constants and $\mu_z$ is a real constant, while in the $(\Gamma_5,\Gamma_6)$ bands for $\mathbb{G}=\mathbf{D}_{3}$ and in the $\Gamma_9$ bands for $\mathbb{G}=\mathbf{D}_{6}$ we have
\begin{equation}
\label{mu-D_3-D_6-final}
    \hat\mu(\bk)=\left(\begin{array}{ccc}
                       \alpha_1(k_x^2-k_y^2) & 2\alpha_2k_xk_y & 0 \\
                       -2\alpha_1k_xk_y & \alpha_2(k_x^2-k_y^2) & 0 \\
                       0 & 0 & \mu_z
                       \end{array}\right),
\end{equation}
where $\alpha_{1,2}$ and $\mu_z$ are real constants, see Appendix \ref{app: D_3} for an example of the calculation. Note that the basal-plane components of $\hat\mu$ vanish at $k_x=k_y=0$ in all four exceptional cases 
for the symmetry reasons. Similar phenomenon has also been found in 3D trigonal and hexagonal centrosymmetric materials, where the transverse components of the effective Zeeman coupling vanish along the main symmetry axis, 
which leads to qualitative changes in the temperature dependence of the spin susceptibility in the superconducting state.\cite{Sam21} The vanishing of the Zeeman coupling perpendicular to the trigonal axis has also 
been reported in bismuth.\cite{Bi-anisotropy}

\subsection{Example: $\mathbb{G}=\mathbf{D}_3$}
\label{sec: one-band-D_3}

This rosette group describes, for instance, the symmetry of a quasi-2D semiconducting quantum well grown along the $[111]$ direction from a bulk semiconductor with diamond or zinc-blende structure,\cite{Winkler-book} or
the surface states of the topological insulator Bi$_2$Te$_3$ (Ref. \onlinecite{Fu09}).
According to Table \ref{table: coreps-point}, there are two types of bands in these materials: the pseudospin bands which correspond to the $\Gamma_4$ corep, and the nonpseudospin bands which correspond 
to the $(\Gamma_5,\Gamma_6)$ corep. 

Assuming that the electron gas is placed in a magnetic field $\bH\perp\hbz$, we obtain from Tables \ref{table: gammas} and \ref{table: ells}, and also Eqs. (\ref{usual-Zeeman}) and (\ref{mu-D_3-D_6-final}), 
the following expressions for the effective one-band Hamiltonians:
\begin{equation}
\label{H-eff-D3-Gamma4}
  \hat{\cal H}=\frac{k^2}{2m^*}+b(H_xk_y-H_yk_x)+(a_1k_y+\mu_\perp H_x)\hat\sigma_1+(-a_1k_x+\mu_\perp H_y)\hat\sigma_2+ia_2(k_+^3-k_-^3)\hat\sigma_3
\end{equation}
in the $\Gamma_4$ bands, and 
\begin{eqnarray}
\label{H-eff-D3-Gamma56}
  \hat{\cal H} &=& \frac{k^2}{2m^*}+b(H_xk_y-H_yk_x)+ia_3(k_+^3-k_-^3)\hat\sigma_3 \nonumber \\
  && + [ia_1(k_+^3-k_-^3)+\alpha_1H_x(k_x^2-k_y^2)-2\alpha_1H_yk_xk_y]\hat\sigma_1 \nonumber \\ 
  && + [a_2(k_+^3+k_-^3)+2\alpha_2H_xk_xk_y+\alpha_2H_y(k_x^2-k_y^2)]\hat\sigma_2
\end{eqnarray}
in the $(\Gamma_5,\Gamma_6)$ bands.

\begin{figure}
  \subfloat{
    \includegraphics[width=.25\textwidth]{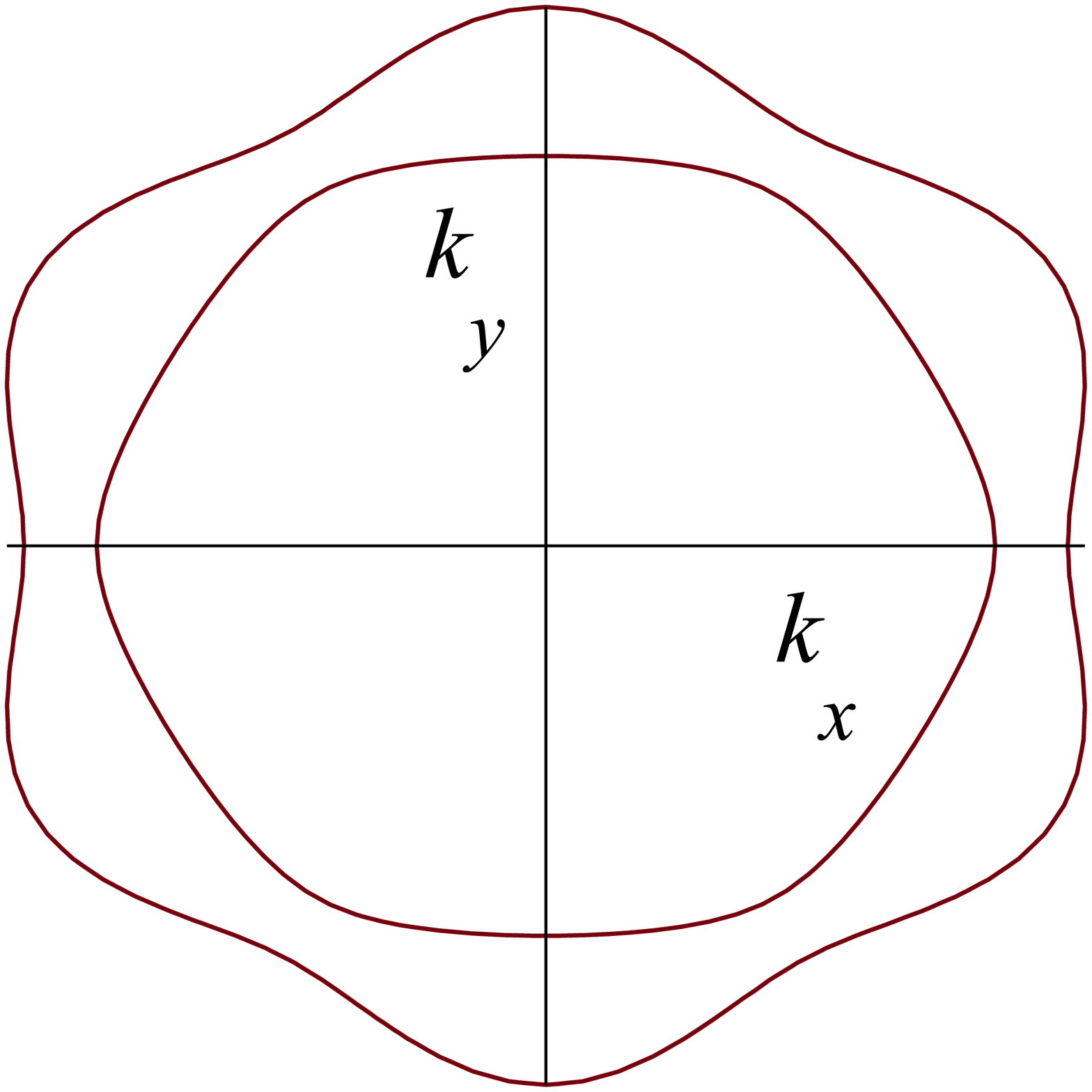}}\\  
  \subfloat{
    \includegraphics[width=.25\textwidth]{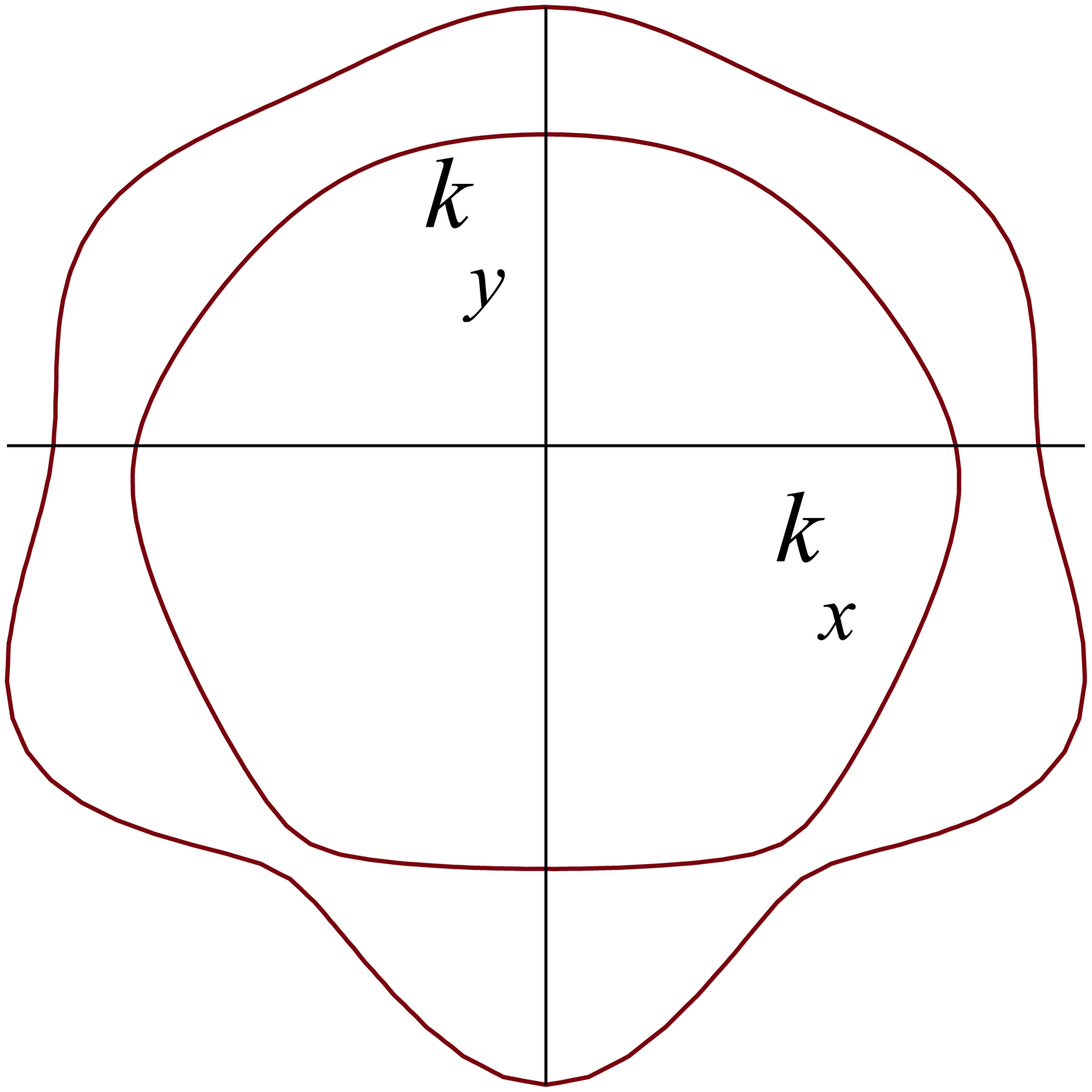}}
\caption{The Fermi surfaces $\xi_\pm(\bk,\bH)=\mu$ ($\mu$ is the chemical potential) of the helicity bands (\ref{D3-bands-G4}), for $\bH=\bm{0}$ (top panel) and for $\bH\neq\bm{0}$ (bottom panel). The Fermi surfaces
in the $(\Gamma_5,\Gamma_6)$ case, see Eq. (\ref{D3-bands-G56}), exhibit a similar behaviour, but the bands remain degenerate at $\bk=\bm{0}$ in the presence of $\bH\parallel\hbx$.}
\label{fig: D3-bands}
\end{figure}

The matrices (\ref{H-eff-D3-Gamma4}) and (\ref{H-eff-D3-Gamma56}) can be easily diagonalized producing two helicity bands $\xi_\pm$, see Eq. (\ref{helicity bands}). Assuming $\bH=H\hbx$, we have
\begin{equation}
\label{D3-bands-G4}
  \xi_\pm(\bk,\bH)=\frac{k^2}{2m^*}+bHk_y\pm\left[a_1^2k_x^2+(a_1k_y+\mu_\perp H)^2+4a_2^2(k_y^3-3k_x^2k_y)^2\right]^{1/2}
\end{equation}
in the $\Gamma_4$ case, and
\begin{eqnarray}
\label{D3-bands-G56}
  \xi_\pm(\bk,\bH) &=& \frac{k^2}{2m^*}+bHk_y\pm\bigl\{[2a_1(k_y^3-3k_x^2k_y)+\alpha_1H(k_x^2-k_y^2)]^2\nonumber\\
  && +[2a_2(k_x^3-3k_y^2k_x)+2\alpha_2Hk_xk_y]^2+4a_3^2(k_y^3-3k_x^2k_y)^2\bigr\}^{1/2}
\end{eqnarray}
in the $(\Gamma_5,\Gamma_6)$ case. In the absence of magnetic field, the helicity bands have a sixfold rotation symmetry, but at $\bH\neq\bm{0}$, this symmetry is lost as the bands get shifted and distorted, 
as shown in Fig. \ref{fig: D3-bands}.
One can see from Eqs. (\ref{D3-bands-G4}) and (\ref{D3-bands-G56}) that there is a qualitative difference between the magnetic response in the $\Gamma_4$ and $(\Gamma_5,\Gamma_6)$ bands. In the former case
the band degeneracy is lifted by the field, whereas in the latter case the bands remain degenerate at the $\Gamma$ point even in a nonzero field, because of the vanishing of the effective Zeeman interaction at $\bk=\bm{0}$.
The degeneracy may be lifted by the terms nonlinear in $\bH$, which we neglected.

\section{Two-band effective Hamiltonian}
\label{sec: two-band H}

The symmetry analysis of the previous section can be straightforwardly extended to the multiband case. Let us consider two twofold degenerate bands corresponding to two $\Gamma$-point coreps, which can be 
the same or different. In the $\Gamma$-point basis $\{|1,1\rangle,|1,2\rangle,|2,1\rangle,|2,2\rangle\}$, see Eq. (\ref{Bloch-states-Gamma-point}), the effective Hamiltonian is given by a $4\times 4$ matrix:
\begin{equation}
\label{H-eff-2-band}
  \hcH(\bk,\bH)=\left( \begin{array}{cc}
                       \hcH_{11}(\bk,\bH) & \hcH_{12}(\bk,\bH) \\
                       \hcH_{21}(\bk,\bH) & \hcH_{22}(\bk,\bH)
                       \end{array} \right),
\end{equation}
where $\hcH_{nn'}$ ($n,n'=1,2$) are $2\times 2$ matrices in the Kramers space satisfying the Hermiticity condition $\hcH_{nn'}(\bk,\bH)=\hcH_{n'n}^\dagger(\bk,\bH)$.
The invariance requirements under the rosette group and TR operations, Eqs. (\ref{H-constraint-g}) and (\ref{H-constraint-K}), hold in the two-band case as well, 
with $\hat{\cal D}$ having the following block-diagonal form:
\begin{equation}
\label{two-band-corep-matrices}
  \hat{\cal D}(g)=\left( \begin{array}{cc}
                     \hat{\cal D}_1(g) & 0 \\
                     0 & \hat{\cal D}_2(g)
                     \end{array} \right),\quad 
  \hat{\cal D}(K)=\left( \begin{array}{cc}
                     -i\hat\sigma_2 & 0 \\
                     0 & -i\hat\sigma_2
                     \end{array} \right),
\end{equation}
which corresponds to a four-dimensional reducible corep. 

The Hamiltonian (\ref{H-eff-2-band}) corresponds to two coupled generalized Rashba models. The intraband ($n=n'$) and interband ($n\neq n'$) components can be represented in the form similar to Eq. (\ref{H-eff-general}):
\begin{equation}
\label{H-eff-2-band-general}
  \hcH_{nn}(\bk,\bH)=\varepsilon_n(\bk,\bH)\hat\sigma_0+\bLam_n(\bk,\bH)\hat{\bm\sigma},\quad \hcH_{12}(\bk,\bH)=\tilde\varepsilon(\bk,\bH)\hat\sigma_0+\tilde\bLam(\bk,\bH)\hat{\bm\sigma},\quad \hcH_{21}=\hcH_{12}^\dagger.
\end{equation}
Here $\varepsilon_{1,2}$ and $\bLam_{1,2}$ are real, but $\tilde\varepsilon$ and $\tilde\bLam$ can be complex.
The symmetry constraints for the intraband components have the same form as Eqs. (\ref{TR-epsilon-Gamma}) and (\ref{g-epsilon-Gamma}), namely,
\begin{equation}
\label{g-intraband-2-band}
  \varepsilon_n(\bk,\bH)=\varepsilon_n(g^{-1}\bk,g^{-1}\bH),\quad \bLam_n(\bk,\bH)=\hat{\cal R}_n(g)\bLam_n(g^{-1}\bk,g^{-1}\bH),
\end{equation}
where the $3\times 3$ orthogonal matrix $\hat{\cal R}_n$ is given by Eq. (\ref{sigma-rotation-nonpseudospin}) in the $n$th band, and
\begin{equation}
\label{TR-intraband-2-band}
  \varepsilon_n(\bk,\bH)=\varepsilon_n(-\bk,-\bH),\quad \bLam_n(\bk,\bH)=-\bLam_n(-\bk,-\bH).
\end{equation}
For the interband components, the constraints become more complicated. From TR invariance, we have
\begin{equation}
\label{TR-interband-2-band}
  \tilde\varepsilon(\bk,\bH)=\tilde\varepsilon^*(-\bk,-\bH),\quad \tilde\bLam_n(\bk,\bH)=-\tilde\bLam_n^*(-\bk,-\bH),
\end{equation}
whereas the rosette group invariance requires that
\begin{equation}
\label{g-interband-2-band}
  \tilde\varepsilon(\bk,\bH)\hat\sigma_0+\tilde\bLam(\bk,\bH)\hat{\bm\sigma}=
  \tilde\varepsilon(g\bk,g\bH)[\hat{\cal D}_1^\dagger(g)\hat{\cal D}_2(g)]+\tilde\bLam(g\bk,g\bH)[\hat{\cal D}_1^\dagger(g)\hat{\bm{\sigma}}\hat{\cal D}_2(g)].
\end{equation}
In particular, if both bands have the same symmetry, i.e., correspond to the same corep, then $\hat{\cal D}_1(g)=\hat{\cal D}_2(g)$ and the last expression takes the same form as Eq. (\ref{g-intraband-2-band}).

\subsection{Example: $\mathbb{G}=\mathbf{D}_{4}$}
\label{sec: D4-2-band}

Due to a large number of possibilities, here we consider only the case of the rosette group $\mathbb{G}=\mathbf{D}_{4}$, which describes, for instance, the symmetry of a conducting interface between two 
insulating oxides LaAlO$_3$ and SrTiO$_3$ (Ref. \onlinecite{interface-SC}). There are two types of bands, 
corresponding to $\Gamma_6$ or $\Gamma_7$ coreps. Using the corep matrices from Table \ref{table: nonpseudospin-corep-matrices}, we obtain that $\hat{\cal R}_n(C_{4z})=\hat R(C_{4z})$ and $\hat{\cal R}_n(\sigma_{y})=\hat R(C_{2y})$, 
therefore, according to Eq. (\ref{g-intraband-2-band}), $\varepsilon_n$ is an invariant scalar and $\bLam_n$ is an invariant pseudovector, for all two-band combinations. 
Expanding $\varepsilon_n$ and $\bLam_n$ in powers of a weak magnetic field, similarly to Eq. (\ref{eps-Gam-expansion}), and using Tables \ref{table: gammas} and \ref{table: ells}, and also Eq. (\ref{usual-Zeeman}),
we arrive at the following expressions for the intraband components of the effective Hamiltonian:
\begin{equation}
\label{D4-intraband}
  \hcH_{nn}(\bk,\bH)=\epsilon_n(\bk)+b_n(H_xk_y-H_yk_x)+(a_nk_y+\mu_{\perp,n}H_x)\hat\sigma_1+(-a_nk_x+\mu_{\perp,n}H_y)\hat\sigma_2,
\end{equation}
for $\bH\perp\hbz$. We assume that
\begin{equation}
\label{epsilons-2-band}
  \epsilon_1(\bk)=\frac{k^2}{2m_1},\quad \epsilon_2(\bk)={\cal E}_b+\frac{k^2}{2m_2},
\end{equation}
where ${\cal E}_b>0$ is the band gap and the effective masses $m_1$ and $m_2$ can be positive or negative, corresponding to electron- or hole-like bands, respectively. 

For the interband components, we have 
\begin{equation}
\label{tildas-expansions}
  \begin{array}{l} 
    \tilde\varepsilon(\bk,\bH)=\tildeps(\bk)+\bH\tilde\bbell(\bk)+O(H^2), \medskip \\ 
    \tilde\Lambda_\nu(\bk,\bH)=\tilde\gamma_\nu(\bk)+\sum\limits_iH_i\tilde\mu_{i\nu}(\bk)+O(H^2),
  \end{array}
\end{equation}
where the $\tilde\bgam$ term will be called the ``interband SO coupling'' (which is not quite accurate, because the SO coupling affects all terms in the effective Hamiltonian). 
It follows from Eq. (\ref{g-interband-2-band}) that $\tilde\varepsilon$ is an invariant scalar and $\tilde\bLam$ is an invariant pseudovector only if both bands correspond to the same corep. In general, we have
\begin{eqnarray}
\label{tilde-epsilon-invariance}
  & \tildeps(\bk)=\tildeps^*(-\bk),\quad \tildeps(\bk)=\pm\tildeps(C_{4z}^{-1}\bk),\quad \tildeps(\bk)=\tildeps(\sigma_y^{-1}\bk),\\
\label{tilde-gamma-invariance}
  & \tilde\bgam(\bk)=-\tilde\bgam^*(-\bk),\quad \tilde\bgam(\bk)=\pm C_{4z}\tilde\bgam(C_{4z}^{-1}\bk),\quad \tilde\bgam(\bk)=C_{2y}\tilde\bgam(\sigma_y^{-1}\bk),\\
\label{tilde-ell-invariance}
  & \tilde\bbell(\bk)=-\tilde\bbell^*(-\bk),\quad \tilde\bbell(\bk)=\pm C_{4z}\tilde\bbell(C_{4z}^{-1}\bk),\quad \tilde\bbell(\bk)=C_{2y}\tilde\bbell(\sigma_y^{-1}\bk),
\end{eqnarray}
and
\begin{equation}
\label{tilde-mu-invariance}
  \hat{\tilde\mu}(\bk)=\hat{\tilde\mu}^*(-\bk),\quad \hat{\tilde\mu}(\bk)=\pm\hat R(C_{4z})\hat{\tilde\mu}(C_{4z}^{-1}\bk)\hat R^{-1}(C_{4z}),\quad
  \hat{\tilde\mu}(\bk)=\hat R(C_{2y})\hat{\tilde\mu}(\sigma_y^{-1}\bk)\hat R^{-1}(C_{2y}),
\end{equation}
instead of Eqs. (\ref{epsilon-invariance}), (\ref{gamma-invariance}), (\ref{ell-invariance}), and (\ref{mu-invariance}), respectively. The upper (lower) sign refers to the bands having the same (different) symmetry. 
This sign difference, combined with the fact that the interband couplings are not necessarily real functions of $\bk$, produces a richer variety of solutions than in the intraband case. 

For instance, one can see from Eq. (\ref{tilde-mu-invariance}) that the interband Zeeman coupling has a simple form
\begin{equation}
\label{tilde-mu-66-77}
  \hat{\tilde\mu}(\bk)=\left(\begin{array}{ccc}
                       \tilde\mu_\perp & 0 & 0 \\
                       0 & \tilde\mu_\perp & 0 \\
                       0 & 0 & \tilde\mu_z
                       \end{array}\right),
\end{equation}
with real $\tilde\mu_\perp$ and $\tilde\mu_z$, only if both bands have the same symmetry. In contrast, if the bands correspond to different coreps, then the interband Zeeman coupling is necessarily momentum-dependent and 
vanishes at $|k_x|=|k_y|$:
\begin{equation}
\label{tilde-mu-67}
  \hat{\tilde\mu}(\bk)=\left(\begin{array}{ccc}
                       \tilde\alpha_\perp(k_x^2-k_y^2) & 0 & 0 \\
                       0 & \tilde\alpha_\perp(k_x^2-k_y^2) & 0 \\
                       0 & 0 & \tilde\alpha_z(k_x^2-k_y^2)
                       \end{array}\right),
\end{equation}
where $\tilde\alpha_\perp$ and $\tilde\alpha_z$ are real constants. Representative expressions for the other interband couplings satisfying the constraints (\ref{tilde-epsilon-invariance}), (\ref{tilde-gamma-invariance}), 
and (\ref{tilde-ell-invariance}) are given in Table \ref{table: two-band Rashba-interband}, see Appendix \ref{app: D_4-interband} for an example of the calculation. 
Note that, unlike the intraband SO coupling, the interband SO coupling is a complex function of $\bk$ and does not have a definite parity. Also note that $\tildeps(\bk)$ cannot contain a constant term, 
even though the constraints (\ref{tilde-epsilon-invariance}) permit a real $\bk$-independent solution in the case of same-symmetry bands. The reason is that such terms cannot appear in the framework of the 
$\bk\cdot\bm{p}$ perturbation theory, see Appendix \ref{app: k p}.

\begin{table}
\caption{The interband couplings of the two-band effective Hamiltonian near the $\Gamma$ point, for $\mathbb{G}=\mathbf{D}_{4}$; $\tilde a_i$ are real constants.}
\begin{tabular}{|c|c|c|c|}
    \hline
	bands		  &  $\tildeps(\bk)$ & $\tilde{\bgam}(\bk)$ & $\tilde\bbell(\bk)$ \\ \hline
    \ $\Gamma_6,\Gamma_6$\ \   &  \ $\tilde a_0(k_x^2+k_y^2)$\ \  &  \ $\tilde a_1(k_y\be_1-k_x\be_2)+i\tilde a_2(k_x^2-k_y^2)k_xk_y\be_3$\ \ &  \ $\tilde a_3(k_y\hbx-k_x\hby)+i\tilde a_4(k_x^2-k_y^2)k_xk_y\hbz$\ \  \\           
    $\Gamma_7,\Gamma_7$   &   &  & \\ \hline                   
    $\Gamma_6,\Gamma_7$   &  $\tilde a_0(k_x^2-k_y^2)$  &  $\tilde a_1(k_y\be_1+k_x\be_2)+i\tilde a_2k_xk_y\be_3$ & $\tilde a_3(k_y\hbx+k_x\hby)+i\tilde a_4k_xk_y\hbz$ \\ \hline                        
\end{tabular}
\label{table: two-band Rashba-interband}
\end{table}

\section{Model of a topological insulator}
\label{sec: TI}

Let us show how a two-band model introduced in Sec. \ref{sec: D4-2-band} can describe a TR invariant topological insulator. At $\bH=\bm{0}$, the effective Hamiltonian (\ref{H-eff-2-band}) becomes
\begin{equation}
\label{H-2-D4-TRI}
  \hcH(\bk)=\left( \begin{array}{cc}
                             \epsilon_1(\bk)\hat\sigma_0+\bgam_1(\bk)\hat{\bm{\sigma}} & \tildeps(\bk)\hat\sigma_0+\tilde{\bgam}(\bk)\hat{\bm{\sigma}} \\
                             \tildeps(\bk)\hat\sigma_0+\tilde{\bgam}^*(\bk)\hat{\bm{\sigma}} & \epsilon_2(\bk)\hat\sigma_0+\bgam_2(\bk)\hat{\bm{\sigma}}
                             \end{array} \right).
\end{equation}
where $\epsilon_{1,2}$ are given by Eq. (\ref{epsilons-2-band}), the intraband SO coupling has the form $\bgam_{1,2}(\bk)=a_{1,2}(k_y\be_1-k_x\be_2)$, see Table \ref{table: gammas}, and the interband couplings depend on 
the band symmetries and can be found in Table \ref{table: two-band Rashba-interband}. In Appendix \ref{app: Upsilon-representation}, we discuss a different (``Dirac'') representation of the two-band Hamiltonian 
using the $4\times 4$ $\Upsilon$ matrices.

The bulk spectrum of the model (\ref{H-2-D4-TRI}) consists of four bands, which are nondegenerate at a general $\bk$, but remain twofold degenerate at $\bk=\bm{0}$ (and other TR invariant momenta) for symmetry reasons, 
see Sec. \ref{sec: Gamma-point}. Due to the TR symmetry, the band energies are even in $\bk$. Indeed, if $\Psi$ is an eigenvector of the matrix (\ref{H-2-D4-TRI}) corresponding to an eigenvalue $E(\bk)$,
then its TR partner $\tilde\Psi=\hat{\cal D}(K)\Psi^*$, with $\hat{\cal D}(K)$ given by Eq. (\ref{two-band-corep-matrices}), is an eigenvector of $\hcH(-\bk)$, 
because $\hcH(-\bk)\tilde\Psi=\hat{\cal D}(K)\hcH^*(\bk)\Psi^*=E(\bk)\tilde\Psi$ [here we used Eq. (\ref{H-constraint-K})]. Thus, $\hcH(\bk)$ and $\hcH(-\bk)$ have the same eigenvalues and one can 
label the bands in such a way that $E_i(\bk)=E_i(-\bk)$, where $i=1,2,3,4$, with $E_1(\bm{0})=E_2(\bm{0})$ and $E_3(\bm{0})=E_4(\bm{0})$.
Explicit analytical solutions for the band energies and the wave functions can only be obtained after some simplifying assumptions, see Sec. \ref{sec: TI-reduced} below. 

The Hamiltonian has the form (\ref{H-2-D4-TRI}) in the $\Gamma$-point basis 
\begin{equation}
\label{basis-A}
  \{|1,1\rangle,|1,2\rangle,|2,1\rangle,|2,2\rangle\},
\end{equation}
where the first index in $|n,s\rangle$ labels the bands, whereas the second one is the Kramers index. To investigate the topological properties of the system, in particular, 
the spectrum of the boundary modes, it is more convenient to switch to a different basis, 
\begin{equation}
\label{basis-B}
  \{|1,1\rangle,|2,2\rangle,|1,2\rangle,|2,1\rangle\},
\end{equation}
in which the Hamiltonian becomes 
\begin{equation}
\label{H-H1-H2}
  \hcH=\frac{{\cal E}_b}{2}+\frac{k^2}{2M}+\hcH_1+\hcH_2,
\end{equation}
where
\begin{equation}
\label{H1H2-2-band}
  \hcH_1(\bk)=
	 \left( \begin{array}{cc}
         \hat h(\bk) & 0 \\
         0 & \hat{\bar h}(\bk)
         \end{array} \right),\quad
  \hcH_2(\bk)=\left( \begin{array}{cc}
         0 & \hat m(\bk) \\
         \hat m^\dagger(\bk) & 0
         \end{array} \right).
\end{equation}
The notations are as follows:
\begin{equation}
\label{h-def}
  \hat h(\bk)=\left( \begin{array}{cc}
         \xi & \tilde\gamma_1-i\tilde\gamma_2 \\
         \tilde\gamma_1+i\tilde\gamma_2 & -\xi
         \end{array} \right),\quad
  \hat{\bar h}(\bk)=\left( \begin{array}{cc}
         \xi & \tilde\gamma_1+i\tilde\gamma_2 \\
         \tilde\gamma_1-i\tilde\gamma_2 & -\xi
         \end{array} \right),
\end{equation}
\begin{equation}
\label{m-def}
  \hat m(\bk)=\left( \begin{array}{cc}
         \gamma_{1,1}-i\gamma_{1,2} & \tildeps+\tilde\gamma_3 \\
         \tildeps+\tilde\gamma_3 & \gamma_{2,1}+i\gamma_{2,2}
         \end{array} \right),
\end{equation}
and
$$
  \xi=\frac{k^2}{2m^*}-\frac{{\cal E}_b}{2},\quad \frac{1}{M}=\frac{1}{2}\left(\frac{1}{m_1}+\frac{1}{m_2}\right),\quad \frac{1}{m^*}=\frac{1}{2}\left(\frac{1}{m_1}-\frac{1}{m_2}\right).
$$
We used the fact that $\tildeps$ and $\tilde\gamma_{1,2}$ are real, whereas $\tilde\gamma_3$ is imaginary, see Table \ref{table: two-band Rashba-interband}. In the basis (\ref{basis-B}), 
the matrix representation of TR is given by $\hat{\cal D}(K)=-i\hat s_2\otimes\hat s_3$, where $\hat{\bm s}$ are the Pauli matrices,\cite{Pauli-matrices} and we obtain from Eq. (\ref{H-constraint-K}) that
\begin{equation}
\label{hp-hm-TR}
  \hat{\bar h}(\bk)=\hat s_3\hat h^*(-\bk)\hat s_3,
\end{equation}
while $\hat m$ satisfies
\begin{equation}
\label{m-TR}
  \hat m(\bk)=-\hat s_3\hat m^\top(-\bk)\hat s_3.
\end{equation}
Thus, the $2\times 2$ Hamiltonians $\hat h$ and $\hat{\bar h}$ are TR partners.

The advantage of the representation (\ref{H-H1-H2}) is that $\hcH_1$ has a block-diagonal structure which can be connected with some well-known models of topological superconductors and insulators. 
The matrices $\hat h$ and $\hat{\bar h}$ have the same form as the Bogoliubov-de Gennes Hamiltonian of a chiral $p$-wave superconductor (for one spin projection), with ${\cal E}_b/2$ playing the role of the chemical potential. 
This system is known to have topologically nontrivial bulk states with gapless boundary modes.\cite{Volovik-book} In the limit $m^*\to\infty$, $\hcH_1$ takes the form of a massive Dirac Hamiltonian 
with the mass given by ${\cal E}_b/2$, and if the band gap changes sign as a function of position, then one arrives at a 2D version of the Volkov-Pankratov model of a topological insulator.\cite{VP85} 
If $m^*$ is finite, then $\hcH_1$ becomes a ``modified'' Dirac Hamiltonian, see Appendix \ref{app: modified Dirac}, and the first three terms in Eq. (\ref{H-H1-H2}) are similar to the Bernevig-Hughes-Zhang model.\cite{BHZ06}

\subsection{Reduced model}
\label{sec: TI-reduced}

Based on the analogy with the models of topological superconductors and insulators, one can expect that our Hamiltonian (\ref{H-H1-H2}) may exhibit nontrivial topology in the bulk, with gapless modes at the boundary. 
To develop an analytically treatable model, we make some simplifying assumptions.
According to Appendix \ref{app: modified Dirac}, $\hcH_1$ describes a topologically nontrivial system only when $m^*>0$. This condition is satisfied if $m_1>0$ and $m_2<0$, which corresponds 
to a ``band inversion'' situation. Furthermore, one can put $m_1=-m_2=m^*$ and $\bgam_1(\bk)=\bgam_2(\bk)=a(k_y\be_1-k_x\be_2)$, without affecting the topological properties of the system. 

We keep the terms of the lowest (linear) order in $\bk$ in $\hcH_2$ and also assume that both bands have the same symmetry, therefore $\tilde{\bgam}(\bk)=\tilde a(k_y\be_1-k_x\be_2)$, 
see Table \ref{table: two-band Rashba-interband}. In this way, dropping the constant ${\cal E}_b/2$, we obtain the following model Hamiltonian in the basis (\ref{basis-B}):
\begin{equation}
\label{model-H-basis B}
  \hcH=\hcH_1+\hcH_2,\quad
  \hcH_1=\left( \begin{array}{cc}
              \hat h & 0 \\
              0 & \hat{\bar h}
              \end{array} \right),\quad
  \hcH_2=\left( \begin{array}{cc}
              0 & \hat m \\
              \hat m^\dagger & 0
              \end{array} \right), 
\end{equation}
where
$$
  \hat h=\xi\hat s_3+\tilde a(k_y\hat s_1-k_x\hat s_2),\quad \hat{\bar h}=\xi\hat s_3+\tilde a(k_y\hat s_1+k_x\hat s_2),\quad \hat m=a(k_y\hat s_0+ik_x\hat s_3),
$$
and $\xi=k^2/2m^*-{\cal E}_b/2$. Without loss of generality, we assume that $a,\tilde a>0$. 
The structure of Eq. (\ref{model-H-basis B}) suggests that, while the interband SO coupling enters the chiral Hamiltonians $\hat h$ and $\hat{\bar h}$ and gives a ``topological character'' to the system, 
the effects of the intraband SO coupling, which are contained in $\hcH_2$, can be treated as a perturbation.

\subsection{Bulk properties; $\mathbb{Z}_2$ invariant}
\label{sec: TI-model-bulk}

\begin{figure}
\includegraphics[width=7cm]{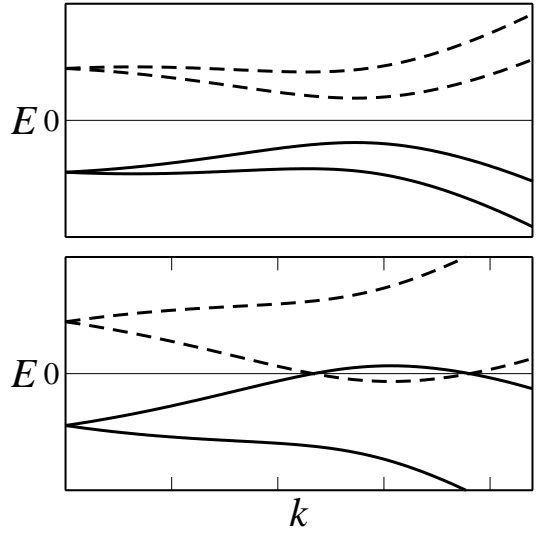}
\caption{Bulk bands of the model (\ref{model-H-basis B}). The solid lines correspond to $E_{1,2}$, and the dashed lines -- to $E_{3,4}$. 
Top panel: a gapped spectrum for $\tilde a>a$ (strong interband SO coupling); bottom panel: a gapless spectrum for $\tilde a<a$ (weak interband SO coupling).}
\label{fig: TI-bands}
\end{figure}

Let us start with the bulk properties of our model. To calculate the bulk spectrum, it is more convenient to return to the basis (\ref{basis-A}), in which the Hamiltonian takes the form 
\begin{equation}
\label{model-H-basis A}
  \hcH=\left( \begin{array}{cc}
              \xi & 0 \\
              0 & -\xi \\
              \end{array} \right)\otimes\hat\sigma_0+
        \left( \begin{array}{cc}
              a & \tilde a \\
              \tilde a & a \\
              \end{array} \right)\otimes\hat q,
\end{equation}
where $\hat q=k_y\hat\sigma_1-k_x\hat\sigma_2$. The $2\times 2$ Kramers blocks in Eq. (\ref{model-H-basis A}) commute and the Hamiltonian can be easily diagonalized. Introducing the eigenstates of $\hat q$:
\begin{equation}
\label{chi-12}
  \chi_+(\bk)=\frac{1}{\sqrt{2}}\left( \begin{array}{c}
              1 \\
              -ie^{i\varphi_{\bk}} \\
              \end{array} \right),\quad 
  \chi_-(\bk)=\frac{1}{\sqrt{2}}\left( \begin{array}{c}
              -ie^{-i\varphi_{\bk}} \\
              1 \\
              \end{array} \right),        
\end{equation}
with $\varphi_{\bk}=\arg(k_x+ik_y)$, which correspond to the eigenvalues $\pm k=\pm\sqrt{k_x^2+k_y^2}$, we obtain the following eigenstates of the Hamiltonian (\ref{model-H-basis A}):
\begin{equation}
\label{Psi-1234}
  \Psi_1=\left( \begin{array}{c}
              w_- \\
              w_+ \\
              \end{array} \right) \otimes \chi_-,\quad 
  \Psi_2=\left( \begin{array}{c}
              w_- \\
              -w_+ \\
              \end{array} \right) \otimes \chi_+,\quad
  \Psi_3=\left( \begin{array}{c}
              w_+ \\
              -w_- \\
              \end{array} \right) \otimes \chi_-,\quad
  \Psi_4=\left( \begin{array}{c}
              w_+ \\
              w_- \\
              \end{array} \right) \otimes \chi_+,
\end{equation}
where
$$
  w_\pm=\frac{1}{\sqrt{2}}\sqrt{1\pm\frac{\xi}{\sqrt{\xi^2+\tilde a^2k^2}}}.
$$
The corresponding eigenvalues form four isotropic bands:
\begin{equation}
\label{TI-bulk-spectrum}
  E_1(\bk)=-E_+,\quad E_2(\bk)=-E_-,\quad E_3(\bk)=E_-,\quad E_4(\bk)=E_+,
\end{equation}
where
$$
  E_\pm(k)=\sqrt{\xi^2(k)+\tilde a^2k^2}\pm ak.
$$
The spectrum qualitatively depends on the relative strength of the intraband and interband SO coupling.
If $a<\tilde a$, then the bulk spectrum is gapped, but if $a>\tilde a$, then the bulk gap closes, see Fig. \ref{fig: TI-bands}. 
We assume that the interband SO coupling is stronger than the intraband one, so that the model describes an insulator with the bands $1$ and $2$ occupied. 
The twofold degeneracy of the bands is lifted by the intraband SO coupling everywhere, except the TR invariant point $\bk=\bm{0}$.

The eigenstates (\ref{Psi-1234}) do not depend on $a$, therefore one can expect that the bulk topology is independent of the strength of the intraband SO coupling (as long as the bulk spectrum remains gapped). 
The Hamiltonian given by Eqs. (\ref{model-H-basis A}) or (\ref{model-H-basis B}) corresponds to the symplectic symmetry class AII and is characterized in 2D by a $\mathbb{Z}_2$ topological invariant.\cite{tenfold-way} 
Among the several equivalent representations of this invariant,\cite{Bernevig-Book} it is the Pfaffian form proposed in Ref. \onlinecite{KM05-2} that is the easiest to calculate for our model. We introduce the ``overlap'' matrix 
$M_{nm}(\bk)=\langle\Psi_n(\bk)|K|\Psi_m(\bk)\rangle$, where $n$ and $m$ label the occupied bands. The overlap matrix is not necessarily unitary, but is antisymmetric at all $\bk$ 
due to the condition $K^2=-1$. In our case, there are two occupied bands, the TR acts in the basis (\ref{basis-A}) as follows: $K\Psi=\mathbb{1}\otimes(-i\hat\sigma_2)\Psi^*$, and, using the the fact that 
$K\chi_+(\bk)=\chi_-(\bk)$ and $K\chi_-(\bk)=-\chi_+(\bk)$, we obtain from Eq. (\ref{Psi-1234}) the following expression for the overlap matrix:
$$
  \hat M(\bk)=\frac{\xi}{\sqrt{\xi^2+\tilde a^2k^2}}
              \left(\begin{array}{cc}
              0 & -1 \\
              1  & 0
              \end{array}\right).
$$
The Pfaffian of this matrix is given by
\begin{equation}
\label{Pfaffian}
  \Pf\hat M(\bk)=-\frac{\xi(\bk)}{\sqrt{\xi^2(\bk)+\tilde a^2k^2}}.
\end{equation}
According to Ref. \onlinecite{KM05-2}, the $\mathbb{Z}_2$ invariant $\nu$ is equal to one half of the number of sign changes of $\Pf\hat M$ along the boundary of a half of the Brillouin zone (HBZ), 
the latter being defined in such a way that it does not contain time-reversed momenta $\bk$ and $-\bk$ at the same time.

In our model, the Brillouin zone extends to the whole momentum plane and one can choose, for instance, the $k_x=0$ line as the HBZ boundary. The zeros of the Pfaffian (\ref{Pfaffian}) are the same as those of $\xi(\bk)$,
namely it vanishes on a circle of radius $\sqrt{m^*{\cal E}_b}$, but only if $m^*$ and ${\cal E}_b$ have the same sign. Thus, we arrive at the following result:
\begin{equation}
\label{Z2-invariant}
  \nu=\left\{ \begin{array}{ll}
                           1,& \mathrm{if}\ m^*{\cal E}_b>0,\medskip \\ 
                           0,& \mathrm{if}\ m^*{\cal E}_b<0.                                       
                           \end{array}\right.
\end{equation}
In the atomic limit, we have $\xi(\bk)=\mathrm{const}$ and the Pfaffian does not change sign, therefore, $\nu=0$ corresponds to a topologically trivial insulator. In the topologically nontrivial case $\nu=1$, the invariant 
(\ref{Z2-invariant}) counts the number of Kramers-degenerate pairs of the boundary zero modes, see the next subsection. Note that $a$ and $\tilde a$ do not explicitly enter Eq. (\ref{Z2-invariant}), therefore 
one can neglect the intraband SO coupling $a$ without affecting the topological properties.

\subsection{Boundary modes}
\label{sec: TI-model-boundary}

Let us now put the model (\ref{model-H-basis B}) in a half-space $x\geq 0$ and assume that the wave functions vanish at $x=0$. The momentum along the boundary is a good quantum number and, making 
the replacement $k_x\to\hat k_x=-i\nabla_x$ in Eq. (\ref{model-H-basis B}), we see that $\hat h$ is the same as $\hat H_+$ from Appendix \ref{app: modified Dirac-boundary}, while $\hat{\bar h}$ is the same as $\hat H_-$, 
with the parameters given by $\alpha=1/2m^*$, $\beta=-{\cal E}_b/2$, and $\gamma=\tilde a$. Without the intraband SO coupling, $\hat h$ and $\hat{\bar h}$ decouple and the Hamiltonian (\ref{model-H-basis B}) 
becomes a modified Dirac Hamiltonian with $m^*{\cal E}_b>0$, which has two counterpropagating linear boundary modes given by
\begin{equation}
\label{TI-boundary-mode}
  E(k_y)=-\tilde ak_y,\quad \bar E(k_y)=\tilde ak_y,
\end{equation}
see Eqs. (\ref{Dirac-E-plus}) and (\ref{Dirac-E-minus}). At $k_y=0$, the two branches cross and form a Kramers pair of the boundary zero modes, whose wave functions have the following form:
\begin{equation}
\label{Psi-0}
  \Psi_0(\br)=\left( \begin{array}{c}
              1 \\
              -1 \\
              0 \\
              0 
              \end{array} \right) F(x),\quad 
  \bar\Psi_0(\br)=\left( \begin{array}{c}
              0 \\
              0 \\
              1 \\
              1 \\
              \end{array} \right) F(x).
\end{equation}
Here $F(x)$ is localized near the boundary, vanishes at $x=0$, and can be chosen to be real, see Eqs. (\ref{Dirac-psi-plus}) and (\ref{Dirac-psi-minus}).

If one takes into account the intraband SO coupling, then it is easy to check that all matrix elements of $\hcH_2(k_y=0)=i\hat s_2\otimes\hat s_3(a\nabla_x)$ in the subspace spanned by the states (\ref{Psi-0}) vanish. 
Therefore, the TR-degenerate boundary zero modes survive at $a\neq 0$, which is consistent with the observation that the variation of $a$ does not affect the bulk topological invariant (\ref{Z2-invariant}).

The boundary zero modes are only destroyed by a TR symmetry-breaking perturbation, for instance, by an external magnetic field. Assuming that the field is parallel to the $xy$ plane, 
so that one does not have to worry about the orbital effects, and keeping the terms of the zeroth order in $\bk$ in the intraband [Eq. (\ref{eps-Gam-expansion})] and interband 
[Eq. (\ref{tildas-expansions})] magnetic couplings, we obtain an additional term in the Hamiltonian (\ref{model-H-basis B}): $\hcH=\hcH_1+\hcH_2+\hcH_3$, where
\begin{equation}
\label{model-H3}
  \hcH_3=\left( \begin{array}{cccc}
              0 & \tilde\mu_\perp H_- & \mu_\perp H_- & 0 \\
              \tilde\mu_\perp H_+ & 0 & 0 & \mu_\perp H_+ \\
              \mu_\perp H_+ & 0 & 0 & \tilde\mu_\perp H_+ \\
              0 & \mu_\perp H_- & \tilde\mu_\perp H_- & 0 \\
              \end{array} \right)
\end{equation}
describes the effective Zeeman coupling in the basis (\ref{basis-B}). Here $H_\pm=H_x\pm iH_y$ and we used Eqs. (\ref{usual-Zeeman}) and (\ref{tilde-mu-66-77}). The matrix elements of Eq. (\ref{model-H3}) in the zero-mode 
subspace have the following form:
$$
  \langle\Psi_0|\hcH_3|\Psi_0\rangle=-\tilde\mu_\perp H_x,\quad \langle\Psi_0|\hcH_3|\bar\Psi_0\rangle=-i\mu_\perp H_y,\quad 
  \langle\bar\Psi_0|\hcH_3|\Psi\rangle=i\mu_\perp H_y,\quad \langle\bar\Psi_0|\hcH_3|\bar\Psi_0\rangle=\tilde\mu_\perp H_x.
$$
Therefore, in the presence of a magnetic field the boundary modes are pushed apart and away from the zero energy, with the energy splitting given by 
$$
  \delta E(k_y=0)=2\sqrt{\tilde\mu_\perp^2H_x^2+\mu_\perp^2H_y^2}.
$$
Note that the field normal to the boundary affects the zero modes only via the interband Zeeman coupling.

\section{Conclusions}
\label{sec: Conclusion}

We derived the phenomenological effective Hamiltonians for 2D electrons in the presence of the SO coupling and a magnetic field, by using the method of invariants. The SO coupling of an arbitrary strength is taken 
into account exactly, by considering the double-valued corepresentations of the magnetic rosette groups. Focusing on the vicinity of the $\Gamma$ point, the effective Hamiltonian is represented 
by an expansion in powers of the electron momentum $\bk$ and the magnetic field $\bH$. In the one-band case, we developed a complete classification of the effective Hamiltonians for all 2D crystal symmetries. 
In the two-band case, due to a large number of possible band combinations, we focused on the example of the rosette group $\mathbf{D}_4$.

We obtained various terms in the effective Hamiltonians representing the intraband and interband antisymmetric SO coupling and the magnetic field interactions. While in some cases these terms reproduce the structure of the 
standard Rashba model, in others, most notably in trigonal and hexagonal crystals, they are significantly different. In the exceptional nonpseudospin bands, namely, $\Gamma_6$ for $\mathbb{G}=\mathbf{C}_{3}$, 
$(\Gamma_{11},\Gamma_{12})$ for $\mathbb{G}=\mathbf{C}_{6}$, $(\Gamma_5,\Gamma_6)$ for $\mathbb{G}=\mathbf{D}_{3}$, and $\Gamma_9$ for $\mathbf{D}_{6}$, all components of the intraband SO coupling are cubic in $\bk$, 
whereas the linear terms are absent for the symmetry reasons, and the in-plane components of the effective Zeeman coupling vanish at the $\Gamma$ point. 

As an application of the two-band effective Hamiltonian for $\mathbb{G}=\mathbf{D}_4$, we introduced a simple model of a TR-invariant topological insulator. In this model, the intraband SO coupling mixes two $2\times 2$ 
TR-partnered chiral Hamiltonians, whose topological character originates from the interband SO coupling. We calculated the spectrum of the boundary modes, which form Kramers pairs protected by a $\mathbb{Z}_2$ 
invariant in the bulk.

\acknowledgments

This work was supported by a Discovery Grant 2021-03705 from the Natural Sciences and Engineering Research Council of Canada.

\appendix

\section{The $\Gamma$-point basis}
\label{app: Bloch bases}

Since all double-valued coreps of the rosette groups are 2D, the Bloch states at the $\Gamma$-point are twofold degenerate. Dropping the band index, these states have the following form:
\begin{equation}
\label{12-states-Gamma-point}
  |1\rangle=\left(\begin{array}{c}
                      \psi_\uparrow(\br) \\
                      \psi_\downarrow(\br)
                      \end{array}\right),\quad
  |2\rangle\equiv K|1\rangle=\left(\begin{array}{c}
                      -\psi^*_\downarrow(\br) \\
                      \psi^*_\uparrow(\br)
                      \end{array}\right),
\end{equation}
where $\psi_\uparrow(\br)$ and $\psi_\downarrow(\br)$ are lattice-periodic functions, which include the effects of the noncentrosymmetric crystal field and the SO coupling. 
Under a symmetry operation $g$, the Bloch states transform as $g|s\rangle=\sum_{s'}|s'\rangle {\cal D}_{s's}(g)$, where $\hat{\cal D}$ is the corep matrix. 
Recall that the action of $g$ on a spinor wave function is given by\cite{Lax-book}
$$
    g\left(\begin{array}{c}
     \psi_\uparrow(\br) \\
     \psi_\downarrow(\br)
     \end{array}\right)=
     \hat D^{(1/2)}(g)\left(\begin{array}{c}
     \psi_\uparrow(g^{-1}\br) \\
     \psi_\downarrow(g^{-1}\br)
     \end{array}\right),
$$
where $\hat D^{(1/2)}(g)$ is the spin-$1/2$ representation of $g$.

Explicit form of the corep matrices can be obtained using the procedure of Ref. \onlinecite{Sam19-1}, with the only modification that the antiunitary element of the symmetry group in our case is the TR operation $K$, 
instead of the conjugation operation $KI$. In the pseudospin bands, the corep matrices are given by $\hat D^{(1/2)}(g)$. The results for the nonpseudospin bands are shown in Table \ref{table: nonpseudospin-corep-matrices}. 
Note that the corep matrices depend on the choice of the basis: a unitary rotation of the basis, $|s\rangle\to|s\rangle'=\sum_{s_1}|s_1\rangle U_{s_1s}$, produces an equivalent corep with 
$\hat{\cal D}'(g)=\hat U^{-1}\hat{\cal D}(g)\hat U$ and $\hat{\cal D}'(K)=\hat U^{-1}\hat{\cal D}(K)\hat U^*$. The ``orientation'' and the phases of the Bloch basis at the $\Gamma$-point can always 
be chosen to reproduce both the matrices in Table \ref{table: nonpseudospin-corep-matrices} and the matrix representation of the TR operation, $\hat{\cal D}(K)=-i\hat\sigma_2$.

Let us consider, for example, a trigonal crystal with $\mathbb{G}=\mathbf{D}_{3}$. This group is generated by the rotation $C_{3z}$ and the reflection $\sigma_{y}$ and has three double-valued irreps: $\Gamma_4$, 
which is 2D and is equivalent to the spin-$1/2$ irrep, and also $\Gamma_5$ and $\Gamma_6$, which are 1D.
The irreps $\Gamma_5$ and $\Gamma_6$ are complex conjugate to each other and pair up to form a single ``physically irreducible'' 2D representation of $\mathbf{D}_{3}$, 
or a 2D Case C corep of the magnetic point group $\mathbf{D}_{3}+K\mathbf{D}_{3}$, which is given by the following matrices:
$$
    \hat{\cal D}(C_{3z})=\left(\begin{array}{cc}
              -1 & 0 \\
              0 & -1
              \end{array}\right),\quad
    \hat{\cal D}(\sigma_{y})=\left(\begin{array}{cc}
              -i & 0 \\
              0 & i
              \end{array}\right),\quad
    \hat{\cal D}(K)=\left(\begin{array}{cc}
              0 & -1 \\
              1 & 0
              \end{array}\right),
$$
where we used the group character tables for $\mathbf{D}_{3}$, see, e.g., Refs. \onlinecite{BC-book} and \onlinecite{Lax-book}. We see that the $(\Gamma_5,\Gamma_6)$ corep is not equivalent to the spin-$1/2$ representation.
It can be brought by a unitary transformation $\hat U=\exp(-i\pi\hat\sigma_1/4)$ to an equivalent form
$$
    \hat{\cal D}(C_{3z})=\left(\begin{array}{cc}
              -1 & 0 \\
              0 & -1
              \end{array}\right),\quad
    \hat{\cal D}(\sigma_{y})=\left(\begin{array}{cc}
              0 & -1 \\
              1 & 0
              \end{array}\right),\quad
    \hat{\cal D}(K)=\left(\begin{array}{cc}
              0 & -1 \\
              1 & 0
              \end{array}\right),
$$
which is the one listed in Table \ref{table: nonpseudospin-corep-matrices}.

\section{Effective Hamiltonians from the $\bk\cdot\bm{p}$ perturbation theory}
\label{app: k p}

In this appendix, we outline how the phenomenological effective Hamiltonian discussed in Sec. \ref{sec: one-band H} can be obtained, at least in principle, from a microscopic theory. 
The eigenfunction of Eq. (\ref{H-general}) which corresponds to the wave vector $\bk$ has the form $\psi_{\bk}(\br)={\cal V}^{-1/2}\varphi_{\bk}(\br)e^{i\bk\br}$, where ${\cal V}$ is the system volume. 
The Bloch factors $\varphi_{\bk}$ are spin-$1/2$ spinors with the same periodicity as the crystal lattice, which are eigenstates of the reduced Hamiltonian
\begin{eqnarray}
\label{H-reduced}
    \hat H_{\bk} &=& \frac{(\hat{\bp}+\hbar\bk)^2}{2m}+U(\br)+\frac{\hbar}{4m^2c^2}\hat{\bm{\sigma}}[\bm{\nabla}U(\br)\times(\hat{\bp}+\hbar\bk)]+\mu_B\bH\hat{\bm\sigma} \nonumber\\
                 &=& \hat H_0+\frac{\hbar^2k^2}{2m}+\delta\hat H,
\end{eqnarray}
where $\hat H_0$ corresponds to $\bk=\bm{0}$ and has the same form as Eq. (\ref{H-general}),
\begin{equation}
\label{h_k}
  \delta\hat H=\hbar\bk\htv+\mu_B\bH\hat{\bm\sigma},
\end{equation}
and $\htv=\hat{\bp}/m+(\hbar/4m^2c^2)(\hat{\bm{\sigma}}\times\bm{\nabla}U)$ is the velocity operator in the presence of the SO coupling. Note that our unperturbed Hamiltonian $\hat H_0$ includes the electron-lattice 
SO coupling, which is assumed to be stronger than the Zeeman interaction $\mu_B\bH\hat{\bm\sigma}$. The latter is included in the perturbation $\delta\hat H$ along with the usual ``$\bk\cdot\bm{p}$'' term. The orbital effects
of the magnetic field are neglected.

In the absence of magnetic field, the energy levels of $\hat H_0$ are twofold degenerate due to the TR symmetry. The spinors $\varphi_{\bk=\bm{0}}(\br)\equiv\langle\br|n,s\rangle$ are labelled by the band index $n$ 
and the Kramers index $s=1,2$, and have the form (\ref{12-states-Gamma-point}). They transform according 
to the irreducible double-valued coreps of the magnetic rosette group (\ref{magnetic G}). The corresponding eigenvalues of $\hat H_0$ are denoted by $\epsilon_n$.

If the complete set of $|n,s\rangle$ is known, then an exact description of the band structure at $\bk\neq\bm{0}$ is provided by diagonalizing the infinite-dimensional matrix $\langle n,s|\hat H_{\bk}|n',s'\rangle$. 
Its elements satisfy certain constraints imposed by the TR symmetry. It follows from $K\htv K^{-1}=-\htv$ and $K\hat{\bm\sigma}K^{-1}=-\hat{\bm\sigma}$ that $K\delta\hat HK^{-1}=-\delta\hat H$.
Therefore, using the antiunitarity of the TR operator, $\langle Ki|j\rangle=\langle i|K^\dagger|j\rangle^*$, and the Hermiticity of $\delta\hat H$, we obtain 
\begin{equation}
\label{matrix-constraint-22-11}
  \langle n,2|\delta\hat H|n',2\rangle=\langle n,1|K^{-1}\delta\hat HK|n',1\rangle^*=-\langle n,1|\delta\hat H|n',1\rangle^* =-\langle n',1|\delta\hat H|n,1\rangle.
\end{equation}
Similarly, using also the fact that $K^2=-1$, we have
\begin{equation}
\label{matrix-constraint-12}
  \langle n,1|\delta\hat H|n',2\rangle=-\langle n,1|(K^{-1}\delta\hat HK)K|n',1\rangle=\langle n,1|K^\dagger\delta\hat H|n',1\rangle=\langle n',1|\delta\hat HK|n,1\rangle=\langle n',1|\delta\hat H|n,2\rangle
\end{equation}  
and
\begin{equation}
\label{matrix-constraint-21}
  \langle n,2|\delta\hat H|n',1\rangle=\langle n',2|\delta\hat H|n,1\rangle.
\end{equation}
The above relations are valid for any $n$ and $n'$. While the diagonal matrix elements $\langle n,s|\delta\hat H|n,s\rangle$ are real, the off-diagonal ones are complex, in general. 

The constraints (\ref{matrix-constraint-22-11}), (\ref{matrix-constraint-12}), and (\ref{matrix-constraint-21}) impose the following matrix structure on the $\bk\cdot\bm{p}$ Hamiltonian:
\begin{equation}
\label{matrix-elements-general}
  \langle n,s|\delta\hat H|n',s'\rangle=iQ_{nn'}\delta_{ss'}+\bm{R}_{nn'}\bm{\sigma}_{ss'},
\end{equation}
where $Q_{nn'}$ and $\bm{R}_{nn'}$ are real quantities satisfying $Q_{nn'}=-Q_{n'n}$ and $\bm{R}_{nn'}=\bm{R}_{n'n}$. Explicitly, 
\begin{eqnarray*}
  && Q_{nn'}=\im\langle n,1|\delta\hat H|n',1\rangle, \\
  && \bm{R}_{nn'}=\bigl(\re\langle n,1|\delta\hat H|n',2\rangle, -\im\langle n,1|\delta\hat H|n',2\rangle, \re\langle n,1|\delta\hat H|n',1\rangle\bigr).
\end{eqnarray*}
Note that $Q_{nn'}$ and $\bm{R}_{nn'}$ are linear in both $\bk$ and $\bH$ and, since the states $|n,s\rangle$ do not have definite parity, $\langle n,s|\htv|n,s'\rangle\neq 0$, in contrast to the centrosymmetric case. 
For illustration, focusing on just two bands, $n,n'=1,2$, we obtain from Eq. (\ref{matrix-elements-general}): 
\begin{equation}
\label{H-matrix-2-bands}
  \hat H_{\bk}=\left( \begin{array}{ccc}
               \epsilon_1+\dfrac{\hbar^2k^2}{2m}+\br_1\hat{\bm{\sigma}} & i\tilde q+\tilde\br\hat{\bm{\sigma}} & \cdots \\      
               -i\tilde q+\tilde\br\hat{\bm{\sigma}} & \epsilon_2+\dfrac{\hbar^2k^2}{2m}+\br_2\hat{\bm{\sigma}} & \cdots \\
               \vdots & \vdots & \ddots 
               \end{array} \right),
\end{equation}
where $\br_1=\bm{R}_{11}$, $\br_2=\bm{R}_{22}$, $\tilde q=Q_{12}$, and $\tilde\br=\bm{R}_{12}$. We use the same notation, $\hat H_{\bk}$, for the reduced Hamiltonian (\ref{H-reduced}) and its matrix representation in 
the $\Gamma$-point basis $\{|1,1\rangle,|1,2\rangle,|2,1\rangle,|2,2\rangle,...\}$.

The effective one-band Hamiltonian can be obtained by eliminating the interband matrix elements $\langle n,s|\hat H_{\bk}|n',s'\rangle$ ($n\neq n'$) in any desired order by a unitary transformation. This procedure,
which is more generally applicable to the matrix elements connecting different groups of degenerate, or quasi-degenerate, states in an arbitrary Hamiltonian,\cite{Lowdin-partitioning,Winkler-book}
is variously known in the literature as the Luttinger-Kohn\cite{LK55} or Schrieffer-Wolff\cite{SW66} transformation and can be traced back to the Foldy-Wouthuysen transformation 
in quantum electrodynamics.\cite{FW50} Let us drop the wave vector subscript and consider a matrix 
\begin{equation}
\label{H-zeta}
  \hat H=\hat H_0+\frac{\hbar^2k^2}{2m}+\zeta\delta\hat H \quad (0\leq\zeta\leq 1),  
\end{equation}
which generalizes Eq. (\ref{H-reduced}). Here $\hat H_0$ is a diagonal matrix with $\langle n,s|\hat H_0|n',s'\rangle=\epsilon_n\delta_{nn'}\delta_{ss'}$ and $\delta\hat H=\delta\hat H'+\delta\hat H''$, with 
the matrices $\delta\hat H'$ and $\delta\hat H''$ having only intraband and interband matrix elements, respectively. For instance, keeping just two bands in Eq. (\ref{H-matrix-2-bands}), we have
\begin{equation}
\label{Hp-Hpp}
  \delta\hat H'=\left( \begin{array}{cc}
               \hat h_1 & 0 \\      
               0 & \hat h_2 
               \end{array} \right),\quad
  \delta\hat H''=\left( \begin{array}{cc}
               0 & \hat{\tilde h} \\      
               \hat{\tilde h}^\dagger & 0
               \end{array} \right),          
\end{equation}
where $\hat h_n=\br_n\hat{\bm{\sigma}}$ and $\hat{\tilde h}=i\tilde q+\tilde\br\hat{\bm{\sigma}}$.
To develop a perturbative expansion in $\delta\hat H$, we introduced a bookkeeping parameter $\zeta$, which will be set to $1$ at the end of the calculation. We assume that the eigenvalues of $\hat H_0$
are well separated from each other, so that the energy splittings between the twofold degenerate bands are much larger than the matrix elements of $\delta\hat H'$ and $\delta\hat H''$.

We shall now try to remove the interband matrix elements from the Hamiltonian (\ref{H-zeta}) by a unitary transformation $\hat H\to\hat H_U=\hat U\hat H\hat U^{-1}$, where $\hat U=\exp(i\hat S)$ and $\hat S$ is a Hermitian matrix, 
which has the same block structure as $\delta\hat H''$, i.e., no intraband elements. Since $\hat S$ vanishes at $\zeta=0$, it can be sought in the form $\hat S=\zeta\hat S_1+\zeta^2\hat S_2+\ldots$, and we obtain:
\begin{equation}
\label{M-prime-expansion}
  \hat H_U=\hat H_0+\frac{\hbar^2k^2}{2m}+\zeta\hat M_1+\zeta^2\hat M_2+O(\zeta^3),
\end{equation}
where
\begin{eqnarray*}
  && \hat M_1=-i[\hat H_0,\hat S_1]+\delta\hat H'+\delta\hat H'',\\
  && \hat M_2=-i[\hat H_0,\hat S_2]-\frac{1}{2}[[\hat H_0,\hat S_1],\hat S_1]-i[\delta\hat H',\hat S_1]-i[\delta\hat H'',\hat S_1].
\end{eqnarray*}
The interband blocks will be removed from $\hat M_1$ if 
\begin{equation}
\label{S_1-equation}
  -i[\hat H_0,\hat S_1]+\delta\hat H''=0,
\end{equation}
from which we obtain $\hat S_1$. Similarly, the interband blocks will be removed from $\hat M_2$ if 
$$
  -i[\hat H_0,\hat S_2]-i[\delta\hat H',\hat S_1]=0,
$$
from which we obtain $\hat S_2$. Repeating this procedure, one can find $\hat S_k$ and $\hat M_k$ by iteration, to any desired order. In particular, $\hat M_1=\delta\hat H'$, 
$\hat M_2=-i[\delta\hat H'',\hat S_1]/2$, etc. Setting $\zeta=1$, we finally obtain the Hamiltonian which contains only intraband terms:
\begin{equation}
\label{H_U}
  \hat H_U=\hat H_0+\frac{\hbar^2k^2}{2m}+\delta\hat H'-\frac{i}{2}[\delta\hat H'',\hat S_1]+\dots,
\end{equation}
where $\hat S_1$ satisfies Eq. (\ref{S_1-equation}). 

For example, neglecting all bands except the two explicitly shown in Eq. (\ref{H-matrix-2-bands}), we seek $\hat S_1$ in the following form:
$$
  \hat S_1=\left( \begin{array}{cc}
               0 & \hat s \\      
               \hat s^\dagger & 0
               \end{array} \right).
$$
Solving the equation (\ref{S_1-equation}) with the interband matrix given by Eq. (\ref{Hp-Hpp}), we find $\hat s=(i/{\cal E}_b)\hat{\tilde h}$, where ${\cal E}_b=\epsilon_2-\epsilon_1$ 
is the band splitting (we assume ${\cal E}_b>0$). Substituting this into Eq. (\ref{H_U}), we obtain: 
\begin{equation}
\label{H_1-effective}
  \hat{\cal H}_1=\epsilon_1+\dfrac{\hbar^2k^2}{2m}+\br_1\hat{\bm{\sigma}}-\frac{1}{{\cal E}_b}(\tilde q^2+\tilde r^2)
\end{equation}
and
\begin{equation}
\label{H_2-effective} 
  \hat{\cal H}_2=\epsilon_2+\dfrac{\hbar^2k^2}{2m}+\br_2\hat{\bm{\sigma}}+\frac{1}{{\cal E}_b}(\tilde q^2+\tilde r^2).
\end{equation}
These $2\times 2$ matrices can be interpreted as the effective one-band Hamiltonians incorporating the corrections of the second order in the interband couplings.
Since $\br_{1,2}$, $\tilde q$, and $\tilde\br$ are linear functions of $\bk$ and $\bH$, one can rewrite the expressions (\ref{H_1-effective}) and (\ref{H_2-effective}) in the form 
\begin{equation}
\label{H-eff-n}
  \hat{\cal H}_n(\bk,\bH)=\epsilon_n(\bk)+\sum_{i\nu}A_{n,i\nu}k_i\hat\sigma_\nu+\sum_{i\nu}B_{n,i\nu}H_i\hat\sigma_\nu+\sum_{ij}C_{n,ij}H_ik_j+O(H^2),
\end{equation}
where we included the quadratic in $\bk$ terms in $\epsilon_n(\bk)$. The coefficients $A$, $B$, and $C$ contain the matrix elements of the velocity $\htv$ and the field $\bH$ in the $\Gamma$-point basis.

Thus, we see that the $\bk\cdot\bm{p}$ perturbation theory combined with the unitary transformation eliminating the interband matrix elements reproduces the general structure of the phenomenological one-band effective Hamiltonian, 
see Eq. (\ref{eps-Gam-expansion}). In particular, the antisymmetric SO coupling corresponds to the second term in Eq. (\ref{H-eff-n}), whereas the last term reproduces the $\bbell$-coupling to the magnetic field. 

We showed in Sec. \ref{sec: one-band H} that in some bands the antisymmetric SO coupling is nonlinear (cubic) in $\bk$. To derive such terms using the $\bk\cdot\bm{p}$ perturbation theory, 
one has include higher orders of the expansion in the interband matrix elements. In order to obtain a two-band effective Hamiltonian of Sec. \ref{sec: two-band H}, one would have to keep the $4\times 4$ matrix 
block shown in Eq. (\ref{H-matrix-2-bands}) and use a unitary transformation to eliminate the interband elements connecting the bands $1$ and $2$ with all other bands.

\section{$\mathbb{G}=\mathbf{D}_{3}$, $(\Gamma_5,\Gamma_6)$ band}
\label{app: D_3}

As an example of calculation of the antisymmetric SO coupling $\bgam(\bk)$ and the generalized Zeeman coupling $\hat\mu(\bk)$ in the exceptional bands, let us consider the $(\Gamma_5,\Gamma_6)$ corep 
in a trigonal crystal with $\mathbb{G}=\mathbf{D}_{3}$. There are two group generators, $g_1=C_{3z}$ and $g_2=\sigma_{y}$, and 
we obtain from Table \ref{table: nonpseudospin-corep-matrices} that $\hat{\cal R}(C_{3z})=\hat{\mathbb 1}$ and $\hat{\cal R}(\sigma_{y})=\hat R(C_{2y})$. Therefore, Eq. (\ref{gamma-invariance}) yields 
the following symmetry constraints:
\begin{equation}
\label{gamma-equations-D_3}
  \bgam(\bk)=\bgam(C_{3z}^{-1}\bk),\quad \bgam(\bk)=C_{2y}\bgam(\sigma_{y}^{-1}\bk).
\end{equation}
Introducing $k_\pm=k_x\pm ik_y$, the first of these constraints becomes $\bgam(k_+,k_-)=\bgam(e^{-2i\pi/3}k_+,e^{2i\pi/3}k_-)$, whose lowest-order odd-degree polynomial solution is given by
$$
  \bgam(\bk)=(b_1k_+^3+b_1^*k_-^3)\be_1+(b_2k_+^3+b_2^*k_-^3)\be_2+(b_3k_+^3+b_3^*k_-^3)\be_3,
$$
where $b_{1,2,3}$ are complex constants. Imposing the second of the constraints (\ref{gamma-equations-D_3}), 
$\gamma_{1,3}(k_+,k_-)=-\gamma_{1,3}(k_-,k_+)$ and $\gamma_2(k_+,k_-)=\gamma_2(k_-,k_+)$, we arrive at the expression in Table \ref{table: gammas}. 

This example makes it clear that the form of $\bgam(\bk)$ depends on the choice of the second generator of a dihedral group, which is in turn determined by the orientation of the coordinate axes relative to the 
crystallographic axes. If one chose $g_2=\sigma_{x}$ (the reflection in the $x=0$ plane) then the second constraint would become $\gamma_1(k_+,k_-)=\gamma_1(-k_-,-k_+)$ and $\gamma_{2,3}(k_+,k_-)=-\gamma_{2,3}(-k_-,-k_+)$, 
therefore
$$
  \bgam(\bk)=ia_1(k_+^3-k_-^3)\be_1+a_2(k_+^3+k_-^3)\be_2+a_3(k_+^3+k_-^3)\be_3,
$$
where $a_{1,2,3}$ are real constants.

From Eq. (\ref{mu-invariance}), the invariance conditions for $\hat\mu$ have the following form: 
\begin{equation}
\label{mu-equations-D_3}
    \hat\mu(\bk)=\hat R(C_{3z})\hat\mu(C_{3z}^{-1}\bk),\quad \hat\mu(\bk)=\hat R(C_{2y})\hat\mu(\sigma_y^{-1}\bk)\hat R^{-1}(C_{2y}), 
\end{equation}
where 
$$
  \hat R(C_{3z})=\left(\begin{array}{ccc}
                       -1/2 & -\sqrt{3}/2 & 0 \\
                       \sqrt{3}/2 & -1/2 & 0 \\
                       0 & 0 & 1  
                       \end{array}\right),\quad 
  \hat R(C_{2y})=\left(\begin{array}{ccc}
                       -1 & 0 & 0 \\
                       0 & 1 & 0 \\
                       0 & 0 & -1  
                       \end{array}\right).
$$
are the rotation matrices. The constraints (\ref{mu-equations-D_3}) have a block-diagonal solution:
\begin{equation}
\label{mu-block-diagonal}
    \hat\mu=\left(\begin{array}{ccc}
                       \mu_{x1} & \mu_{x2} & 0 \\
                       \mu_{y1} & \mu_{y2} & 0 \\
                       0 & 0 & \mu_{z3}
                       \end{array}\right).
\end{equation}
While the equation for $\mu_{z3}$ is trivially satisfied by a constant $\mu_{z3}(\bk)=\mu_z$, the solutions for the other components are more involved. Introducing 
$u^\pm_1=\mu_{x1}\pm i\mu_{y1}$ and $u^\pm_2=\mu_{x2}\pm i\mu_{y2}$, we obtain:
$u^\pm_1(\bk)=e^{\pm 2\pi i/3}u^\pm_1(C_{3z}^{-1}\bk)$ and $u^\pm_2(\bk)=e^{\pm 2\pi i/3}u^\pm_2(C_{3z}^{-1}\bk)$.
One can easily see that the basal-plane components of the effective Zeeman coupling necessarily depend on $\bk$ and vanish at $\bk=\bm{0}$.
Using the lowest-order even-degree polynomial solutions for $u^\pm_{1,2}(\bk)$, we have
$$
    \hat\mu(\bk)=\left(\begin{array}{ccc}
                       \beta_1k_+^2+\beta_1^*k_-^2 & \beta_2k_+^2+\beta_2^*k_-^2 & 0 \\
                       i\beta_1k_+^2-i\beta_1^*k_-^2 & i\beta_2k_+^2-i\beta_2^*k_-^2 & 0 \\
                       0 & 0 & \mu_z
                       \end{array}\right),
$$
where $\beta_{1,2}$ are complex constants and $\mu_z$ is a real constant. Imposing now the second of the constraints (\ref{mu-equations-D_3}), we obtain Eq. (\ref{mu-D_3-D_6-final}).

\section{Interband couplings for $\mathbb{G}=\mathbf{D}_{4}$}
\label{app: D_4-interband}

To illustrate the calculations of the interband couplings in a two-band Rashba model (\ref{H-eff-2-band}) for $\mathbb{G}=\mathbf{D}_{4}$, 
we consider $\tildeps(\bk)$ and $\tilde\bgam(\bk)$ in the case when the bands have different symmetry, i.e., when one band corresponds to the $\Gamma_6$ corep and the other -- to the $\Gamma_7$ corep. 
We obtain from Eq. (\ref{tilde-epsilon-invariance}) that $\re\tildeps$ is even in $\bk$, whereas $\im\tildeps$ is odd. On the other hand, the invariance 
under $C_{2z}$ means that $\tildeps(\bk)=\tildeps(-\bk)$, therefore $\tildeps(\bk)$ is real. It follows from Eq. (\ref{tilde-epsilon-invariance}) that
$\tildeps(k_+,k_-)=-\tildeps(-ik_+,ik_-)$ and $\tildeps(k_+,k_-)=\tildeps(k_-,k_+)$. 
The lowest-order even-degree polynomial solution of these equations has the form $\tildeps\propto k_+^2+k_-^2$, therefore $\tildeps(\bk)\propto k_x^2-k_y^2$.

From Eq. (\ref{tilde-gamma-invariance}), we obtain that $\re\tilde\bgam$ is odd in $\bk$, whereas $\im\tilde\bgam$ is even. The invariance under $C_{2z}$ means that $\tilde\gamma_{1,2}(\bk)=-\tilde\gamma_{1,2}(-\bk)$ 
and $\tilde\gamma_3(\bk)=\tilde\gamma_3(-\bk)$, therefore $\tilde\gamma_1$ and $\tilde\gamma_2$ are real, but $\tilde\gamma_3$ is purely imaginary. Introducing $\tilde\gamma_\pm=\tilde\gamma_1\pm i\tilde\gamma_2$, 
the invariance conditions under $C_{4z}$ become
$\tilde\gamma_\pm(k_+,k_-)=\mp i\tilde\gamma_\pm(-ik_+,ik_-)$ and $\tilde\gamma_3(k_+,k_-)=-\tilde\gamma_3(-ik_+,ik_-)$.
The lowest-order polynomial solution of these equations has the following form: $\tilde\gamma_\pm=c_\pm k_\mp$ and $\tilde\gamma_3=c_1k_+^2+c_2k_-^2$. 
The invariance under $\sigma_y$ imposes additional constraints: $\tilde\gamma_\pm(k_+,k_-)=-\tilde\gamma_\mp(k_-,k_+)$ and $\tilde\gamma_3(k_+,k_-)=-\tilde\gamma_3(k_-,k_+)$.
Therefore, $c_-=-c_+$, $c_2=-c_1$, and we arrive at the expressions in Table \ref{table: two-band Rashba-interband}.

\section{Two-band Hamiltonian in the $\Upsilon$-matrix representation}
\label{app: Upsilon-representation}

Any effective two-band Hamiltonian, see Eq. (\ref{H-eff-2-band}), can be represented as a linear combination of 16 Hermitian $4\times 4$ matrices as follows:
\begin{equation}
\label{H-2-band-linear-combination}
  \hat{\cal H}(\bk,\bH)=\sum_{p,q=0}^3\omega_{pq}(\bk,\bH)\hat\tau_p\otimes\hat\sigma_q,
\end{equation}
where $\hat\tau_0$ and $\hat\tau_{1,2,3}$ ($\hat\sigma_0$ and $\hat\sigma_{1,2,3}$) are the identity matrix and the Pauli matrices acting on the band (Kramers) indices and the coefficients $\omega_{pq}$ are real. 
The TR invariance constraint (\ref{H-constraint-K}), with the matrices $\hat{\cal D}(K)$ given by Eq. (\ref{two-band-corep-matrices}), takes the form
\begin{equation}
\label{H-2-band-TR-constraint}
  \sum_{p,q=0}^3\omega_{pq}(\bk,\bH)\hat\tau_p\otimes\hat\sigma_q=\sum_{p,q=0}^3\omega_{pq}(-\bk,-\bH)\hat U_K^\dagger(\hat\tau_p\otimes\hat\sigma_q)^*\hat U_K,
\end{equation}
where $\hat U_K=\hat\tau_0\otimes i\hat\sigma_2$. It is straightforward to check that six of the matrices $\hat\tau_p\otimes\hat\sigma_q$ are invariant under TR, in the sense that
$\hat U_K^\dagger(\hat\tau_p\otimes\hat\sigma_q)^*\hat U_K=\hat\tau_p\otimes\hat\sigma_q$, whereas the remaining ten change sign. The six TR-even matrices are the unit matrix $\hat\tau_0\otimes\hat\sigma_0\equiv\hat\Upsilon_0$ 
and the following five traceless matrices:
\begin{equation}
\label{Clifford-Gamma-matrices}
  \hat\tau_1\otimes\hat\sigma_0\equiv\hat\Upsilon_1,\quad \hat\tau_3\otimes\hat\sigma_0\equiv\hat\Upsilon_2,\quad \hat\tau_2\otimes\hat\sigma_1\equiv\hat\Upsilon_3,\quad \hat\tau_2\otimes\hat\sigma_2\equiv\hat\Upsilon_4,\quad
  \hat\tau_2\otimes\hat\sigma_3\equiv\hat\Upsilon_5.
\end{equation}
The matrices $\hat{\Upsilon}_a$ ($a=1,...,5$) satisfy the relations
\begin{equation}
\label{Clifford-algebra}
  \{\hat\Upsilon_a,\hat\Upsilon_b\}=2\delta_{ab},
\end{equation}
i.e., generate a Clifford algebra. The ten TR-odd matrices can be chosen as
$$
  \hat\Upsilon_{ab}=\frac{1}{2i}[\hat\Upsilon_a,\hat\Upsilon_b]=-i\hat\Upsilon_a\hat\Upsilon_b,\quad a<b.
$$
It is easy to see that $\hat\Upsilon_a^2=\hat\Upsilon_{ab}^2=\hat\Upsilon_0$. Note that we use the $\Gamma$-point basis $\{|1,1\rangle,|1,2\rangle,|2,1\rangle,|2,2\rangle\}$ and our nomenclature for the $\Upsilon$ matrices 
follows Ref. \onlinecite{KM05-2}.

In terms of the $\Upsilon$ matrices, Eq. (\ref{H-2-band-linear-combination}) takes the form
\begin{equation}
\label{H-2-band-Gamma-representation}
  \hat{\cal H}(\bk,\bH)=d_0(\bk,\bH)\hat\Upsilon_0+\sum_{a}d_a(\bk,\bH)\hat\Upsilon_a+\sum_{a<b}d_{ab}(\bk,\bH)\hat\Upsilon_{ab}.
\end{equation}
The expansion coefficients here are real and satisfy the following constraints: 
\begin{equation}
\label{d_a-constraints}
  d_0(-\bk,-\bH)=d_0(\bk,\bH),\quad d_a(-\bk,-\bH)=d_a(\bk,\bH),\quad d_{ab}(-\bk,-\bH)=-d_{ab}(\bk,\bH).
\end{equation}
which follow from the TR invariance requirement (\ref{H-2-band-TR-constraint}). Further constraints are imposed by the point-group symmetries. 
For example, in the case of $\mathbb{G}=\mathbf{D}_4$ in zero field we obtain from Eq. (\ref{H-2-D4-TRI}) and Table \ref{table: two-band Rashba-interband} that only the following coefficients are nonzero:
\begin{eqnarray*}
  && d_0=\frac{\epsilon_1+\epsilon_2}{2},\quad d_1=\tildeps,\quad d_2=\frac{\epsilon_1-\epsilon_2}{2},\quad d_5=i\tilde\gamma_3, \\
  && d_{13}=\frac{a_1-a_2}{2}k_y,\quad d_{14}=-\frac{a_1-a_2}{2}k_x,\quad d_{23}=-\tilde\gamma_1,\quad d_{24}=-\tilde\gamma_2,\quad d_{35}=\frac{a_1+a_2}{2}k_x,\quad d_{45}=\frac{a_1+a_2}{2}k_y.
\end{eqnarray*}
The reduced two-band model of a topological insulator introduced in Sec. \ref{sec: TI-reduced}, corresponds to
\begin{eqnarray*}
  && d_0=\frac{{\cal E}_b}{2},\quad d_2=\xi=\frac{k^2}{2m^*}-\frac{{\cal E}_b}{2},\\ 
  && d_{23}=-\tilde ak_y,\quad d_{24}=\tilde ak_x,\quad d_{35}=ak_x,\quad d_{45}=ak_y,
\end{eqnarray*}
with all other coefficients equal to zero.  

The spectrum of the Hamiltonian (\ref{H-2-band-Gamma-representation}) has a simple analytical form in the ``Clifford limit'', when Eq. (\ref{H-2-band-Gamma-representation}) 
contains only the matrices forming a Clifford algebra, in addition to the unit matrix. To illustrate this, let us consider a \textit{centrosymmetric} TR-invariant crystal, in which case 
the coreps and the bands have a definite parity. Setting $\bH=\bm{0}$ and $g=I$ ($I$ is the spatial inversion), the constraint (\ref{H-constraint-g}) takes the form 
\begin{equation}
\label{H-2-constraint-I}
  \hcH(\bk)=\hat U_I^\dagger\hcH(-\bk)\hat U_I,
\end{equation}
where
$$
  \hat U_I=\hat{\cal D}(I)=\left(\begin{array}{cc}
                  p_1 & 0 \\
                  0 & p_2
                  \end{array}\right)\otimes\hat\sigma_0,
$$
and $p_{1,2}$ are the band parities. A straightforward calculation shows that $\hat\Upsilon_0$ and $\hat\Upsilon_2$ are always inversion-even, in the sense that $\hat U_I^\dagger\hat\Upsilon_{0,2}\hat U_I=\hat\Upsilon_{0,2}$. The 
other $\Upsilon$ matrices can be even or odd, depending on the relative parity of the bands, i.e. $\hat U_I^\dagger\hat\Upsilon_a\hat U_I=p_1p_2\hat\Upsilon_a$, for $a=1,3,4,5$. Similarly, $\hat\Upsilon_{ab}$ is inversion-even 
if $a,b\neq 2$, otherwise $\hat U_I^\dagger\hat\Upsilon_{ab}\hat U_I=p_1p_2\hat\Upsilon_{ab}$. 

From Eq. (\ref{H-2-constraint-I}) and the TR constraints (\ref{d_a-constraints}), we obtain that if the bands have the same parity, then all $d_{ab}=0$ and the Hamiltonian becomes
\begin{equation}
\label{Clifford-same-parity}
  \hat{\cal H}(\bk)=d_0(\bk)\hat\Upsilon_0+\sum_{a=1}^5 d_a(\bk)\hat\Upsilon_a,
\end{equation}
i.e., it contains only the Clifford algebra matrices and the unit matrix. This last expression can be easily diagonalized by squaring the second term and using the Clifford algebra relations 
(\ref{Clifford-algebra}). In this way, we find that the bands are twofold degenerate and given by
\begin{equation}
\label{Clifford-spectrum-same}
  E_\pm(\bk)=d_0(\bk)\pm\sqrt{\sum_a d_a^2(\bk)}.
\end{equation}
If the bands have opposite parity, then $d_a=0$ ($a\neq 2$) and $d_{ab}=0$ ($a,b\neq 2$), therefore
\begin{equation}
\label{Clifford-opposite-parity}
  \hat{\cal H}(\bk)=d_0(\bk)\hat\Upsilon_0+d_2(\bk)\hat\Upsilon_2+d_{12}(\bk)\hat\Upsilon_{12}+d_{23}(\bk)\hat\Upsilon_{23}+d_{24}(\bk)\hat\Upsilon_{24}+d_{25}(\bk)\hat\Upsilon_{25}.
\end{equation}
Since the five matrices $\hat\Upsilon_2$, $\hat\Upsilon_{12}$, $\hat\Upsilon_{23}$, $\hat\Upsilon_{24}$, and $\hat\Upsilon_{25}$ form a Clifford algebra, the spectrum consists of two twofold degenerate bands:
\begin{equation}
\label{Clifford-spectrum-opposite}
  E_\pm(\bk)=d_0(\bk)\pm\sqrt{d_2^2(\bk)+d_{12}^2(\bk)+d_{23}^2(\bk)+d_{24}^2(\bk)+d_{25}^2(\bk)}.
\end{equation}
The expressions (\ref{Clifford-spectrum-same}) and (\ref{Clifford-spectrum-opposite}) reproduce, in a rather circuitous way, the well-known statement that the Bloch bands in a centrosymmetric TR-invariant crystal are twofold 
degenerate at each $\bk$. Note that, although $\hcH(\bk)\neq\hcH(-\bk)$ if the bands have opposite parity, see Eq. (\ref{Clifford-opposite-parity}), the band dispersions are always even in $\bk$.

\section{Modified Dirac Hamiltonian}
\label{app: modified Dirac}

In this Appendix, we review the properties of the modified Dirac Hamiltonian of the form $\hcH=\diag(\hat H_+,\hat H_-)$, with
\begin{equation}
\label{Dirac-Hpm}
  \hat H_\pm(\bk)=\left( \begin{array}{cc}
         \alpha(k_x^2+k_y^2)+\beta & \gamma(k_y\pm ik_x) \\
         \gamma(k_y\mp ik_x) & -\alpha(k_x^2+k_y^2)-\beta,
         \end{array} \right)=\bm{g}_\pm(\bk)\hat{\bm s},
\end{equation}
where $\hat{\bm s}$ are the Pauli matrices and $\bm{g}_\pm=(\gamma k_y,\mp \gamma k_x,\alpha k^2+\beta)$. Note that $\hat H_-(\bk)=\hat s_3\hat H_+^*(-\bk)\hat s_3$, which can be interpreted as a consequence of
TR invariance, according to Eq. (\ref{hp-hm-TR}). While the $2\times 2$ blocks $\hat H_\pm$ have the same bulk spectrum consisting of two symmetric gapped branches, $\sqrt{\gamma^2k^2+(\alpha k^2+\beta)^2}$ 
and $-\sqrt{\gamma^2k^2+(\alpha k^2+\beta)^2}$, their opposite chiralities manifest themselves in different spectra of the topologically protected gapless boundary modes.

\subsection{Topology in the bulk}
\label{app: modified Dirac-bulk}

In order to construct a bulk topological invariant for our 2D system, we introduce, following Ref. \onlinecite{Volovik-book}, an auxiliary real variable $\omega$ and define the Green's function for 
$\hat H=\hat H_+$ or $\hat H_-$ as follows: $\hat G(\bk,\omega)=[i\omega-\hat H(\bk)]^{-1}$. Then, the topological invariant has the form
\begin{equation}
\label{bulk-N}
  N=-\frac{1}{24\pi^2}\int\mathrm{Tr}(\hat Gd\hat G^{-1})^3,
\end{equation}
where the integration is performed over $\omega$ and $\bk=(k_x,k_y)$. For $\hat H(\bk)=\bm{g}(\bk)\hat{\bm s}$, we obtain: 
\begin{equation}
\label{Dirac-N3}
  N=\frac{1}{4\pi}\int d^2\bk\, \hat{\bm{g}}\left(\frac{\partial\hat{\bm{g}}}{\partial k_x}\times\frac{\partial\hat{\bm{g}}}{\partial k_y}\right),
\end{equation}
where $\hat{\bm{g}}=\bm{g}/|\bm{g}|$. 

Since at $|\bk|\to\infty$, $\hat{\bm{g}}_\pm\to(0,0,\sign\,\alpha)=\mathrm{const}$, the 2D momentum space is isomorphic to the $S^2$ sphere and $N_\pm$ is an integer (the winding number) equal to the degree of a mapping
$\bk\to\hat{\bm{g}}_\pm(\bk)$, i.e. $S^2\to S^2$. After a straightforward calculation, Eq. (\ref{Dirac-N3}) yields the following result:
\begin{equation}
\label{Dirac-N3-final}
  N_\pm=\pm\frac{1}{2}(\sign\,\beta-\sign\,\alpha),
\end{equation}
therefore $\hat H_+$ and $\hat H_-$ describe a topologically nontrivial bulk system if $\alpha\beta<0$, then $|N_\pm|=1$. In contrast, if $\alpha\beta>0$, then $N_\pm=0$ and the system is topologically trivial. This conclusion 
will be confirmed by an explicit calculation of the spectrum of the boundary modes in Sec. \ref{app: modified Dirac-boundary} below. Also, one can show that the first Chern number, which is given 
by the Berry flux through the Brillouin zone, is equal to the winding number (\ref{Dirac-N3}). For this reason, the Hamiltonians $\hat H_+$ and $\hat H_-$ describe what is sometimes called the Chern insulators.\cite{Bernevig-Book} 

The topological invariant (\ref{bulk-N}) can also be defined for the full modified Dirac Hamiltonian $\hcH=\diag(\hat H_+,\hat H_-)$ as follows:
${\cal N}=-(1/24\pi^2)\int\mathrm{Tr}(\hat{\cal G}d\hat{\cal G}^{-1})^3$, where $\hat{\cal G}(\bk,\omega)=[i\omega-\hcH(\bk)]^{-1}$. It is easy to see that the total Chern number ${\cal N}=N_++N_-=0$, 
according to Eq. (\ref{Dirac-N3-final}). This does not mean, however, that the system is topologically trivial, since one can define another invariant:
\begin{equation}
\label{Dirac-nu-invariant}
  \nu=\frac{1}{2}|N_+-N_-|=\left\{ \begin{array}{ll}
                           1,\ \mathrm{if}\ \alpha\beta<0,\\
                           0,\ \mathrm{if}\ \alpha\beta>0.                                       
                           \end{array}\right.
\end{equation}
This is the $\mathbb{Z}_2$ invariant counting the number of time-reversed pairs of counterpropagating boundary modes, see the next subsection.

\subsection{Boundary modes}
\label{app: modified Dirac-boundary}

Let us consider a half-plane $x\geq 0$ and replace $\bk\to\hat\bk=-i\bm{\nabla}$ in Eq. (\ref{Dirac-Hpm}). The momentum $k_y$ is a good quantum number and the eigenstates of $\hat H_\pm$ have the form $e^{ik_yy}\psi(x)$. 
We assume that the wave functions vanish at $x=0$ (a ``hard wall'' boundary condition). At given $k_y$, the Hamiltonians become
\begin{equation}
\label{Dirac-H-pm-operators}
  \hat H_\pm(k_y)=\left( \begin{array}{cc}
         \alpha(k_y^2-\nabla_x^2)+\beta & \gamma(k_y\pm\nabla_x) \\
         \gamma(k_y\mp\nabla_x) & -\alpha(k_y^2-\nabla_x^2)-\beta,
         \end{array} \right),
\end{equation}
whose solutions corresponding to the energy $E$ and localized near the surface have the form $\psi(x)=(u,v)^\top e^{-\kappa x}$, where $\re\kappa>0$. For $\hat H_+$, we find
\begin{equation}
\label{Dirac-uv}
  \frac{u}{v}=\frac{\gamma(k_y-\kappa)}{E+\alpha z-\beta}=\frac{E-\alpha z+\beta}{\gamma(k_y+\kappa)},
\end{equation}
where $z=\kappa^2-k_y^2$. The last expression produces an equation for $\kappa$, which has two solutions, $\kappa_{1,2}^2=k_y^2+z_{1,2}$, with
\begin{equation}
\label{Dirac-z1z2}
  z_{1,2}=d\pm\sqrt{d^2-\frac{\beta^2-E^2}{\alpha^2}},\quad d=\frac{\beta}{\alpha}+\frac{\gamma^2}{2\alpha^2}.
\end{equation}
Therefore, at given $k_y$, the general localized eigenstate has the following form:
\begin{equation}
\label{Dirac-general solution}
  \psi(x)=C_1\left(\begin{array}{c}
                u_1 \\ v_1
                \end{array}\right)
                e^{-\kappa_1 x}
          +C_2\left(\begin{array}{c}
                u_2 \\ v_2
                \end{array}\right)
                e^{-\kappa_2 x}.
\end{equation}
Note that at $\alpha=0$, i.e., in the ``Dirac limit'', there is only one localized solution and the hard wall boundary condition is not applicable, because it would lead to $\psi(x)=0$ at all $x\geq 0$. 

Substituting the solution (\ref{Dirac-general solution}) into the boundary condition $\psi(0)=0$ and using Eq. (\ref{Dirac-uv}), we arrive at the following two equivalent equations for $E$:
$$
  E=-\beta-\alpha(k_y+\kappa_1)(k_y+\kappa_2),\quad E=\beta+\alpha(k_y-\kappa_1)(k_y-\kappa_2).
$$
Adding and subtracting these equations, we obtain:
\begin{equation}
\label{Dirac-E-eq1}
  E=-\alpha(\kappa_1+\kappa_2)k_y,
\end{equation}
which means that $E$ is an odd function of $k_y$, and
\begin{equation}
\label{Dirac-E-eq2}
  \kappa_1\kappa_2=-\frac{\beta}{\alpha}-k_y^2.
\end{equation}
Squaring Eq. (\ref{Dirac-E-eq1}) and taking into account Eq. (\ref{Dirac-z1z2}), we have $E^2=\gamma^2k_y^2$, therefore the boundary mode energy has the form $E(k_y)=\zeta|\gamma|k_y$, where the coefficient $\zeta$ 
satisfies $\zeta^2=1$, does not depend on $k_y$, and can be calculated at $k_y=0$. 

Setting $k_y=0$ in Eq. (\ref{Dirac-E-eq2}), we see that the localized solution with $\re\kappa_{1,2}>0$ exists only if the signs of $\alpha$ and $\beta$ are opposite, which agrees with the bulk topological argument given above. 
Focusing on the topological regime with $\alpha\beta<0$ and introducing a dimensionless parameter 
$$
  \varrho=\frac{\gamma^2}{2|\alpha\beta|}-1,
$$
we obtain from Eq. (\ref{Dirac-z1z2}):
\begin{equation}
\label{Dirac-kappas}
  \kappa_1(k_y=0)=\sqrt{\frac{|\beta|}{|\alpha|}}f^{1/2}(\varrho),\quad \kappa_2(k_y=0)=\sqrt{\frac{|\beta|}{|\alpha|}}f^{-1/2}(\varrho),
\end{equation}
where $f(\varrho)=\varrho+\sqrt{\varrho^2-1}$. If $\rho>1$, then $f$, $\kappa_1$, and $\kappa_2$ are all real positive, and it follows from Eq. (\ref{Dirac-E-eq1}) that $\zeta=-\sign(\alpha)$.
If $|\varrho|<1$, then $f=e^{i\phi}$ ($0<\phi<\pi$), $\kappa_2=\kappa_1^*$, with $\re\kappa_{1,2}>0$, and we again obtain $\zeta=-\sign(\alpha)$.
Thus, we finally arrive at the following exact expression for the energy of a single nondegenerate boundary mode of the Hamiltonian $\hat H_+$:
\begin{equation}
\label{Dirac-E-plus}
  E_+(k_y)=-\sign(\alpha)|\gamma|k_y=\sign(\beta)|\gamma|k_y.
\end{equation}
Similarly, for $\hat H_-$ we obtain:
\begin{equation}
\label{Dirac-E-minus}
  E_-(k_y)=\sign(\alpha)|\gamma|k_y=-\sign(\beta)|\gamma|k_y.
\end{equation}
Thus, at $\alpha\beta<0$ the modified Dirac Hamiltonian $\hcH$ has two counterpropagating chiral boundary modes, given by Eqs. (\ref{Dirac-E-plus}) and (\ref{Dirac-E-minus}). 
It is easy to see that these modes can be obtained from one another by TR. Indeed, the Hamiltonians (\ref{Dirac-H-pm-operators}) satisfy $\hat H_-(k_y)=\hat s_3\hat H_+^*(-k_y)\hat s_3$. 
If $\psi_{+,k_y}(x)$ is an eigenstate of $\hat H_+$ corresponding to the eigenvalue $E_+(k_y)$, then its time-reversed partner $\psi_{-,k_y}(x)=\hat s_3\psi_{+,-k_y}^*(x)$ is an eigenstate of $\hat H_-$ 
corresponding to the eigenvalue $E_-(k_y)=E_+(-k_y)$.

Regarding the explicit expressions for the boundary mode wave functions, we focus on $k_y=0$ and obtain from Eqs. (\ref{Dirac-uv}) and (\ref{Dirac-kappas}) that the zero mode for $\hat H_+$ has the form
\begin{equation}
\label{Dirac-psi-plus}
  \psi_+(x)=C\left(\begin{array}{c}
                   1 \\ -\sign(\alpha\gamma)
                   \end{array}\right)
	    \left(e^{-\kappa_1x}-e^{-\kappa_2x}\right).
\end{equation}
For $\hat H_-$, we have
\begin{equation}
\label{Dirac-psi-minus}
  \psi_-(x)=C\left(\begin{array}{c}
                   1 \\ \sign(\alpha\gamma)
                   \end{array}\right)
	    \left(e^{-\kappa_1x}-e^{-\kappa_2x}\right).
\end{equation}
If we choose the zero-mode wave functions to be real, then the normalization coefficient at $\varrho>1$ is given by $C=\sqrt{\kappa_1\kappa_2(\kappa_1+\kappa_2)}/(\kappa_1-\kappa_2)$. 
At $|\varrho|<1$, one can write $\kappa_{1,2}=\kappa\pm iq$ ($\kappa$, $q$ are real positive), and $C=i\sqrt{\kappa(\kappa^2+q^2)/2q^2}$. The zero modes (\ref{Dirac-psi-plus}) and (\ref{Dirac-psi-minus}) 
are TR partners: $\psi_-(x)=\hat s_3\psi_+^*(x)$. It is also worth noting that the coordinate dependence of the boundary mode qualitatively changes as $\varrho$ varies, from a superposition of two exponentials at $\varrho>1$ 
to a damped sinusoid at $|\varrho|<1$.


\begin{thebibliography}{99}
\bibitem{Pauli-matrices}
We use different notations for the Pauli matrices depending on the context: while $\hat{\bm s}$ are the general Pauli matrices, $\hat{\bm\sigma}$ act on the Kramers indices, and
$\hat{\bm\tau}$ act on the band indices.
\end{thebibliography}
\end{document}